\documentclass[%
superscriptaddress,
frontmatterverbose,
preprint,
showpacs,
preprintnumbers,
 amsmath,amssymb,
 aps,
 prc,
]{revtex4-1}
\usepackage{graphicx}
\usepackage{dcolumn}
\usepackage{bm}
\usepackage{color}
\usepackage{CJK}
\def\ve#1{{\bm{#1}}}
\def\nuc#1#2#3{{}^{#2}_{#3}\mathrm{#1}}
\def\urm#1{\scriptstyle{\text{\textrm{\textmd{\textup{#1}}}}}}
\def\uurm#1{\scriptscriptstyle{\text{\textrm{\textmd{\textup{#1}}}}}}
\def\Nabla{\bm{\nabla}}
\let\temp\epsilon
\let\epsilon\varepsilon
\let\varepsilon\temp
\let\temp\relax
\begin{document}
%
\begin{CJK*}{UTF8}{}
\preprint{RIKEN-QHP-382}
\preprint{RIKEN-iTHEMS-Report-18}
\title{Coulomb exchange functional with generalized gradient approximation for self-consistent Skyrme Hartree-Fock calculations}
\author{Tomoya Naito (\CJKfamily{min}{内藤智也})}
\affiliation{Department of Physics, Graduate School of Science, The University of Tokyo,
  Tokyo 113-0033, Japan}
\affiliation{RIKEN Nishina Center, Wako 351-0198, Japan}
\author{Xavier Roca-Maza}
\affiliation{Dipartimento di Fisica, Universit\`{a} degli Studi di Milano,
  Via Celoria 16, 20133 Milano, Italy}
\affiliation{INFN, Sezione di Milano,
  Via Celoria 16, 20133 Milano, Italy}
\author{Gianluca Col\`{o}}
\affiliation{Dipartimento di Fisica, Universit\`{a} degli Studi di Milano,
  Via Celoria 16, 20133 Milano, Italy}
\affiliation{INFN, Sezione di Milano,
  Via Celoria 16, 20133 Milano, Italy}
\author{Haozhao Liang (\CJKfamily{gbsn}{梁豪兆})}
\email{haozhao.liang@riken.jp}
\affiliation{RIKEN Nishina Center, Wako 351-0198, Japan}
\affiliation{Department of Physics, Graduate School of Science, The University of Tokyo,
  Tokyo 113-0033, Japan}
\date{\today}
\begin{abstract}
  We perform self-consistent Skyrme Hartree-Fock calculations with the Coulomb exchange functional using the generalized gradient approximation (GGA).
  It is found that the Perdew-Burke-Ernzerhof GGA (PBE-GGA) Coulomb exchange functional is able to reproduce the exact-Fock energy for nuclei in a wide region of the nuclear chart with one adjustable parameter. 
  The remaining error of the GGA Coulomb exchange energy with respect to the exact-Fock energy dominantly comes from the functional-driven error.
\end{abstract}
\maketitle
\end{CJK*}
%
\section{Introduction}
\par
Atomic nuclei are composed of protons and neutrons that interact with one another through the nuclear and electromagnetic forces.
Since it is much stronger than the electromagnetic force,
the nuclear force mainly determines the properties of atomic nuclei.
Nevertheless, in specific studies it is important to evaluate the electromagnetic contribution to the properties of atomic nuclei,
for example, for
the mass difference of mirror nuclei \cite{Nolen1969Annu.Rev.Nucl.Sci.19_471},
the energy of the isobaric analog state \cite{Jaenecke1965Nucl.Phys.73_97,Shlomo1978Rep.Prog.Phys.41_957,Roca-Maza2018Phys.Rev.Lett.120_202501},
the isospin symmetry breaking terms of the nuclear force \cite{Coon1982Phys.Rev.C26_2402},
and the superallowed Fermi $ \beta $-decay \cite{Liang2009Phys.Rev.C79_064316,Hardy2015Phys.Rev.C91_025501}.
Since the static electromagnetic force is well known and mainly associated with the Coulomb contribution,
in principle it is possible to evaluate the contribution of electromagnetic force for such phenomena with high accuracy.
\par
The exchange term of a two-body interaction is characteristic for fermionic systems.
In nuclear physics, the Coulomb exchange term is calculated in the exact form in some studies, including nonrelativistic \cite{Titin-Schnaider1974Phys.Lett.B49_397,Anguiano2001Nucl.Phys.A683_227,Skalski2001Phys.Rev.C63_024312,LeBloas2011Phys.Rev.C84_014310,Roca-Maza2016Phys.Rev.C94_044313}
and relativistic Hartree-Fock calculations \cite{Long2006Phys.Lett.B640_150,Liang2008Phys.Rev.Lett.101_122502}.
However, due to the numerical cost, the Coulomb exchange energy density functional is usually treated within 
the local density approximation (LDA)
(i.e.,~the Hartree-Fock-Slater approximation \cite{Dirac1930Proc.Camb.Phil.Soc.26_376,Slater1951Phys.Rev.81_385}
and its relativistic version \cite{Gu2013Phys.Rev.C87_041301,Shen2016Chin.Phys.Lett.33_102103,Shen2017Phys.Rev.C96_014316}),
or even neglected \cite{Bender2003Rev.Mod.Phys.75_121,Liang2015Phys.Rep.570_1}.
\par
Recently it has been shown that Coulomb energy density functionals built by using the generalized gradient approximation (GGA) give almost the same accuracy for the total energy as the exact-Fock calculation \cite{Naito2018Phys.Rev.C97_044319}, using the experimental charge density distribution as the input of the functional.
As a step further, the corresponding self-consistent calculations performed within the Skyrme Hartree-Fock theory,
as well as a quantitative discussion of the results, are highly desired. 
Self-consistent calculations with the GGA Coulomb exchange functional have an advantage 
since the numerical cost of the self-consistent calculation with the GGA Coulomb exchange functional is $ O \left( N^3 \right) $,
while that with the exact-Fock term is $ O \left( N^4 \right) $ \cite{Engel2011_Springer-Verlag}.
\par
One of the relevant issues is the free parameter $ \mu $ that appears in the Perdew-Burke-Ernzerhof GGA (PBE-GGA) Coulomb exchange functional,
which will be defined in Sec.~\ref{sec:theo}.
As we discuss in detail below, the form of the PBE-GGA functional was determined in order to satisfy several physical conditions \cite{Perdew1996Phys.Rev.B54_16533,Perdew1996Phys.Rev.Lett.77_3865}.
Two different values of $ \mu $ have been widely used in the studies of atoms \cite{Perdew1996Phys.Rev.Lett.77_3865} and solids \cite{Perdew2008Phys.Rev.Lett.100_136406}, respectively.
In this paper, we carry out self-consistent Skyrme Hartree-Fock calculations by using the PBE-GGA functional instead of the exact-Fock term.
Therefore, the optimal value of $ \mu $ and the applicability of our choice will be discussed in detail.   
\par
This paper is organized in the following way:
First, the theoretical framework and general discussion for the PBE-GGA is given in Sec.~\ref{sec:theo}.
Second, the calculation setup is explained in Sec.~\ref{sec:calc}.
The systematic calculations are shown and discussed in Sub.~\ref{subsec:sys} and a detailed analysis for $ \nuc{Pb}{208}{} $ is provided in Sub.~\ref{subsec:detail}.
Finally, the conclusions and perspectives of this work are shown in Sec.~\ref{sec:conc}.
The form of the potential and the rearrangement term for the GGA Coulomb exchange functional are shown in the appendix.
%
\section{Theoretical Framework}
\label{sec:theo}
\par
The LDA Coulomb exchange functional \cite{Dirac1930Proc.Camb.Phil.Soc.26_376,Slater1951Phys.Rev.81_385} is well known under 
the name of Hartree-Fock-Slater approximation and reads
\begin{equation}
  \label{eq:x-LDA}
  E_{\urm{Cx}}^{\urm{LDA}}
  \left[
    \rho_{\urm{ch}}
  \right]
  =
  - \frac{3}{4}
  \frac{e^2}{4 \pi \epsilon_0}
  \left( \frac{3}{\pi} \right)^{1/3}
  \int
  \left[
    \rho_{\urm{ch}} \left( \ve{r} \right)
  \right]^{4/3}
  \, d \ve{r} ,
\end{equation}
where
$ \rho_{\urm{ch}} $ is the charge density distribution.
To go beyond the LDA, the GGA Coulomb exchange functionals \cite{Perdew1992Phys.Rev.B46_6671,Perdew1996Phys.Rev.Lett.77_3865,Perdew2008Phys.Rev.Lett.100_136406} have been proposed as 
\begin{equation}
  \label{eq:x-PBE}
  E_{\urm{Cx}}^{\urm{PBE}}
  \left[
    \rho_{\urm{ch}}
  \right]
  =
  - \frac{3}{4}
  \frac{e^2}{4 \pi \epsilon_0}
  \left( \frac{3}{\pi} \right)^{1/3}
  \int
  \left[
    \rho_{\urm{ch}} \left( \ve{r} \right)
  \right]^{4/3}
  F \left(s \left( \ve{r} \right) \right)
  \, d \ve{r},
\end{equation}
where $ F $ is the enhancement factor due to the density gradient.
Here, $ s $ denotes the dimensionless density gradient
\begin{equation}
  \label{eq:s}
  s = \frac{\left| \Nabla \rho_{\urm{ch}} \right|}{2 k_{\urm{F}} \rho_{\urm{ch}}},
  \qquad
  k_{\urm{F}}
  =
  \left(
    3 \pi^2 \rho_{\urm{ch}}
  \right)^{1/3}.
\end{equation}
\par
In particular, the enhancement factor $ F $ in the PBE-GGA Coulomb exchange functional is assumed to be \cite{Perdew1996Phys.Rev.Lett.77_3865}
\begin{equation}
  \label{eq:F-PBE}
  F \left( s \right)
  =
  1 + \kappa
  -
  \frac{\kappa}{1 + \mu s^2 / \kappa},
\end{equation}
in order to satisfy some physical conditions shown below \cite{Perdew1996Phys.Rev.B54_16533}.
Accordingly, the parameters $ \kappa $ and $ \mu $ are determined to satisfy the same conditions.
\par
First, in the uniform density distribution, i.e.,~$ s = 0 $, the PBE-GGA Coulomb exchange functional should correspond to the LDA one.
Thus,
\begin{equation}
  \label{eq:uniform}
  F \left( 0 \right) = 1 
\end{equation}
is trivially required.
\par
Second, the Coulomb exchange functional should satisfy the Lieb-Oxford bound \cite{Lieb1981Int.J.QuantumChem.19_427}, that is an analytical inequality
\begin{equation}
  \label{eq:Lieb_Oxford}
  E_{\urm{Cx}} \left[ \rho_{\urm{ch}} \right]
  \ge
  -1.679
  \int
  \left[ \rho_{\urm{ch}} \left( \ve{r} \right) \right]^{4/3}
  \, d \ve{r},
\end{equation}
which is derived from the H\"{o}lder inequality in mathematics \cite{Lieb2001_AmericanMathematicalSociety}.
To satisfy this condition, the parameter $ \kappa $ is determined as $ \kappa = 0.804 $
for any value of $ \mu $.
\par
Third, the PBE-GGA Coulomb exchange functional should also satisfy the uniform scaling \cite{Levy1985Phys.Rev.A32_2010}
\begin{equation}
  \label{eq:scaling}
  E_{\urm{Cx}}
  \left[ \zeta^3 \rho_{\urm{ch}} \left( \zeta \ve{r} \right) \right]
  =
  \zeta E_{\urm{Cx}}
  \left[ \rho_{\urm{ch}} \left( \ve{r} \right) \right], 
\end{equation}
for any $ \zeta $, in order to satisfy the condition of the exchange hole
\begin{equation}
  \label{eq:integ_P}
  \int
  P_{\parallel} \left( \ve{r}, \ve{r} + \ve{u} \right)
  \, d \ve{u}
  =
  -1
\end{equation}
for all $ \ve{r} $, where 
\begin{align}
  P_{\parallel} \left( \ve{r}_1, \ve{r}_2 \right)
  & =
    \frac{1}{N \left(N - 1 \right)}
    \sum_{\sigma = \uparrow, \downarrow}
    \rho^{\sigma} \left( \ve{r}_1 \right)
    \left[
    \rho^{\sigma} \left( \ve{r}_2 \right)
    +
    \rho_{\urm{x}}^{\sigma} \left( \ve{r}_1, \ve{r}_2 \right)
    \right], \label{eq:P_para} \\
  \rho^{\sigma} \left( \ve{r} \right)
  & =
    \sum_j
    \left| \psi_{j \sigma} \left( \ve{r} \right) \right|^2, \\
  \rho_{\urm{x}}^{\sigma} \left( \ve{r}_1, \ve{r}_2 \right)
  & =
    -
    \frac{\left| \sum_j
    \psi^*_{j \sigma} \left( \ve{r}_1 \right) \,
    \psi_{j \sigma} \left( \ve{r}_2 \right) \right|^2}
    {\rho^{\sigma} \left( \ve{r}_1 \right)}.
\end{align}
Here, $ \psi_{j \sigma} $ is the Kohn-Sham single-particle orbital of the $ j $th occupied state,
$ \sigma $ is the spin coordinate, whereas $ \sigma = \uparrow $ and $ \sigma = \downarrow $ represent the spin-up and spin-down states, respectively.
The quantity defined in Eq.~\eqref{eq:P_para} is the normalized pair-density probability of 
finding simultaneously two fermions at $ \ve{r}_1 $ and $ \ve{r}_2 $ with spin $ \sigma $.
Consequently, the physical meaning of Eq.~\eqref{eq:integ_P} is related to the Pauli blocking \cite{Engel2011_Springer-Verlag,Martin2004_CambridgeUniversityPress}.
\par
Finally, at the slowly varying limit, i.e.,~$ s \simeq 0 $,
in order to recover the LDA response function,
the PBE-GGA Coulomb functional represents the linear response of the homogeneous electron gas as \cite{Perdew1996Phys.Rev.B54_16533}
\begin{equation}
  \label{eq:slowly_1}
  \lim_{s \to 0}
  F
  \left( s \right)
  =
  1 + \mu s^2,
\end{equation}
with $ \mu = 0.21951 $.
This $ \mu $ can be understood as a coefficient in the response function theory since it multiplies a term proportional to the square of the gradient of the charge density \cite{Perdew1996Phys.Rev.B54_16533,Antoniewicz1985Phys.Rev.B31_6779}.
\par
Combining Eqs.~\eqref{eq:uniform}--\eqref{eq:scaling}, and \eqref{eq:slowly_1},
the PBE-GGA enhancement factor $ F $ is determined as in Eq.~\eqref{eq:F-PBE}.
The latter condition \eqref{eq:slowly_1} is not unique. 
The enhancement factor of the PBEsol-GGA Coulomb exchange functional \cite{Perdew2008Phys.Rev.Lett.100_136406} is also determined to satisfy the same conditions as the PBE-GGA one,
while in the limit $ s \to 0 $ it holds $ \mu = 0.1235 $ instead \cite{Antoniewicz1985Phys.Rev.B31_6779}.
It is empirically known that the PBEsol-GGA Coulomb functional reproduces the electron structure of solids better than the PBE-GGA functional.
\par
This discussion can be applied to proton systems if protons are assumed to be point particles
since protons and electrons share common properties from the point of view of electromagnetic interaction.
However, the value of $ \mu $ cannot be uniquely determined.
Therefore, the coefficient $ \mu $ can be considered as a free parameter,
while the coefficient $ \kappa $ is fixed.
\section{Calculation Setup}
\label{sec:calc}
\par
The LDA and PBE-GGA Coulomb exchange functionals are used in the self-consistent Skyrme Hartree-Fock calculations \cite{Vautherin1972Phys.Rev.C5_626},
while for the nuclear part of the EDF the SAMi parameter set \cite{Roca-Maza2012Phys.Rev.C86_031306} is used.
Here, one should note that no refit of the nuclear part of the EDF is needed due to the use of the PBE-GGA functional for the Coulomb exchange term 
since this term produces at most a difference of $ 1 \, \mathrm{MeV} $ in the total binding energy with respect to LDA.
Therefore, the difference does not deteriorate the quality of SAMi in the description of bulk nuclear 
properties such as binding energies or charge radii.
In any case, the purpose of the present work is to demonstrate the reliability and accuracy of the proposed local form to calculate the Coulomb exchange energy and,
therefore, we will not concentrate on the comparison with experiment.
In this paper protons are treated as point particles as is usually done, i.e.,~$ \rho_{\urm{ch}} = \rho_p $,
where $ \rho_{\urm{ch}} $ and $ \rho_p $ are the ground-state charge and proton density distributions, respectively.
Since the PBE-GGA Coulomb exchange functional is written in terms of the density,
the form factor of nucleon could be considered in the self-consistent steps, and this finite-size effect will be considered as a future study.
\par
For the numerical calculation, the \textsc{skyrme\_rpa} code \cite{Colo2013Comput.Phys.Commun.184_142} is used.
In this code, spherical symmetry is assumed and for the present calculations,
a box of $ 15 \, \mathrm{fm} $ with a mesh of $ 0.1 \, \mathrm{fm} $ is used.
There is no need of pairing correlations for the doubly-magic nuclei, 
while the pairing correlations are neglected in open-shell nuclei.
Although the pairing correlations might be important for a detailed comparison to experiment, this is not the aim of the present study.
The purpose here is to understand if our new local Coulomb functional can provide a satisfactory description of the exact Coulomb exchange energy in the Skyrme Hartree-Fock calculations for finite nuclei when calculated within the same conditions.
\par
In the Skyrme Hartree-Fock calculations, the total energy can be written in two ways.
One way is
\begin{equation}
  \label{eq:energy_gs1}
  E_{\urm{gs}}
  =
  T_0
  +
  E_{\urm{nucl}} \left[ \rho_p, \rho_n \right]
  +
  E_{\urm{Cd}} \left[ \rho_{\urm{ch}} \right]
  +
  E_{\urm{Cx}} \left[ \rho_{\urm{ch}} \right],
\end{equation}
where $ \rho_n $ is the ground-state neutron density distribution,
$ T_0 $ is the kinetic energy, and 
$ E_{\urm{nucl}} $, $ E_{\urm{Cd}} $, and $ E_{\urm{Cx}} $ are nuclear, Coulomb direct, and Coulomb exchange functionals, respectively.
The other way is 
\begin{equation}
  \label{eq:energy_gs2}
  E_{\urm{gs}}
  =
  \frac{1}{2}
  \sum_j
  \left(
    \epsilon_j
    +
    \tau_j
  \right)
  +
  E_{\urm{rea}},
\end{equation}
where $ \epsilon_j $ and $ \tau_j $ are the single-particle energy and single-particle kinetic energy, respectively,
and $ E_{\urm{rea}} $ is the energy of the rearrangement term.
In all of the present calculations, we find $ \left( E_{\urm{tot}}^2 - E_{\urm{tot}}^1 \right) / E_{\urm{tot}}^1 $ is of the order of $ 10^{-6} $,
where $ E_{\urm{tot}}^1 $ and $ E_{\urm{tot}}^2 $ are the total energies calculated by Eqs.~\eqref{eq:energy_gs1} and \eqref{eq:energy_gs2}, respectively.
Hence, the present numerical accuracy is satisfactory \cite{Roca-Maza2018Prog.Part.Nucl.Phys.101_96}.
\section{Results and Discussion}
\label{sec:res}
\subsection{Systematic calculations}
\label{subsec:sys}
\par
\begin{figure}[t]
  \centering
  \includegraphics[width=8cm]{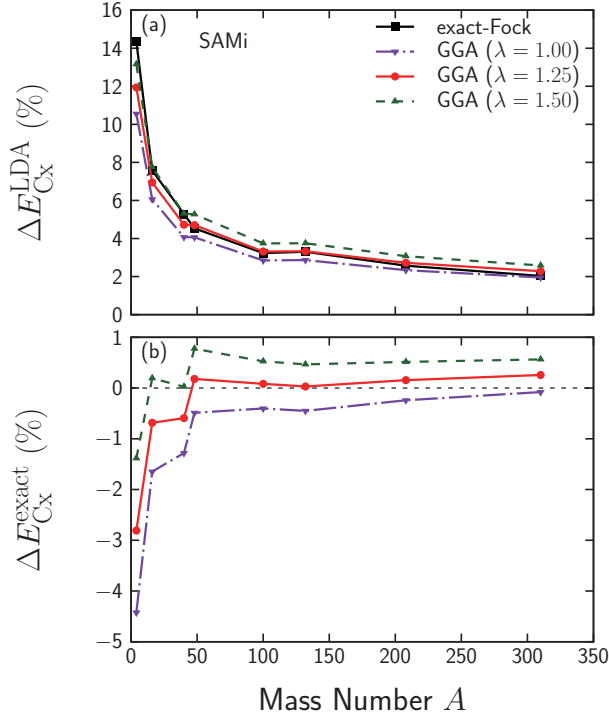}
  \caption{
    (a) The deviation of the Coulomb exchange energy $ E_{\urm{Cx}} $ obtained with PBE-GGA from that obtained within LDA, $ \Delta E_{\urm{Cx}}^{\urm{LDA}} $; 
    (b) the same deviation but with respect to exact-Fock, $ \Delta E_{\urm{Cx}}^{\urm{exact}} $. 
    Both quantities are displayed as a function of $ A $ for doubly magic nuclei.
    For the nuclear part of the EDF, the SAMi functional \cite{Roca-Maza2012Phys.Rev.C86_031306} is used.
    Results calculated using PBE-GGA with $ \lambda = 1.00 $, $ 1.25 $, and $ 1.50 $ are shown with down triangles, circles, and up triangles, respectively.
    For comparison, the exact-Fock results are shown with squares.}
  \label{fig:systematic_dmagic}
\end{figure}

The Coulomb exchange energies $ E_{\urm{Cx}} $ for the doubly-magic nuclei calculated with the LDA and PBE-GGA Coulomb exchange functionals are shown in Table \ref{tab:sys_Ecx_dmagic}.
For comparison, the exact-Fock energies are also calculated,
by using first-order perturbation theory \cite{Roca-Maza2016Phys.Rev.C94_044313}.
This is assumed to be accurate enough for the purpose of the present discussion. 
To see the difference between the LDA and PBE-GGA clearly,
the deviation of the Coulomb exchange energy $ E_{\urm{Cx}} $ of PBE-GGA from that of LDA, $ \Delta E_{\urm{Cx}}^{\urm{LDA}} $, 
and the deviation from that of exact-Fock $ \Delta E_{\urm{Cx}}^{\urm{exact}} $; that is,
\begin{equation}
  \label{eq:ecx_diff}
  \Delta E_{\urm{Cx}}^{\urm{LDA}}
  =
  \frac{E_{\urm{Cx}} - E_{\urm{Cx}}^{\urm{LDA}}}{E_{\urm{Cx}}},
  \qquad
  \Delta E_{\urm{Cx}}^{\urm{exact}}
  =
  \frac{E_{\urm{Cx}} - E_{\urm{Cx}}^{\urm{exact}}}{E_{\urm{Cx}}},
\end{equation}
are shown as a function of mass number $ A $ in Figs.~\ref{fig:systematic_dmagic}(a) and \ref{fig:systematic_dmagic}(b), respectively.
Results calculated from the exact-Fock and PBE-GGA are shown with squares and down triangles, respectively.
It is seen that in the light-mass region $ \Delta E_{\urm{Cx}}^{\urm{LDA}} $ is larger than $ 10 \, \% $, while in the medium-heavy- and heavy-mass regions $ \Delta E_{\urm{Cx}}^{\urm{LDA}} $ decreases gradually with $ A $.
This is because the ratio of the surface region to the volume region in the light nuclei is larger than that in the medium-heavy or heavy nuclei, as discussed in Ref.~\cite{Naito2018Phys.Rev.C97_044319}.
From light nuclei to heavy nuclei, the PBE-GGA results show similar behavior as the exact-Fock results.
\par
However, it is also seen that the absolute values of the PBE-GGA Coulomb exchange energies are slightly smaller than those of the exact-Fock energy, systematically.
To improve this, the free parameter of the PBE-GGA Coulomb exchange functional, $ \mu $, is multiplied by a factor $ \lambda $.
According to Eq.~\eqref{eq:slowly_1}, larger values of $ \lambda $ give larger enhancement factors $ F $.
The Coulomb exchange energies $ E_{\urm{Cx}} $ calculated with $ \lambda = 1.25 $ and $ 1.50 $ are shown in Table \ref{tab:sys_Ecx_dmagic},
while the corresponding deviations $ \Delta E_{\urm{Cx}}^{\urm{LDA}} $ and $ \Delta E_{\urm{Cx}}^{\urm{exact}} $ defined by Eq.~\eqref{eq:ecx_diff} are shown with circles and up triangles, respectively, in Fig.~\ref{fig:systematic_dmagic}.
It is found that, in the light-mass region, in order to reproduce the exact-Fock results, $ \lambda = 1.50 $ or more is required,
while in the medium-heavy- and heavy-mass regions $ \lambda = 1.25 $ reproduces well the exact-Fock results.
The PBE-GGA result with $ \lambda = 1.00 $ reproduces the exact-Fock result in the case of the superheavy nucleus $ \nuc{126}{310}{} $.
The behavior in Fig.~\ref{fig:systematic_dmagic} has also been determined for other Skyrme functionals (e.g.,~SLy5), showing a negligible dependence on the parametrization used.
\par
\begin{figure}[t]
  \centering
  \includegraphics[width=8cm]{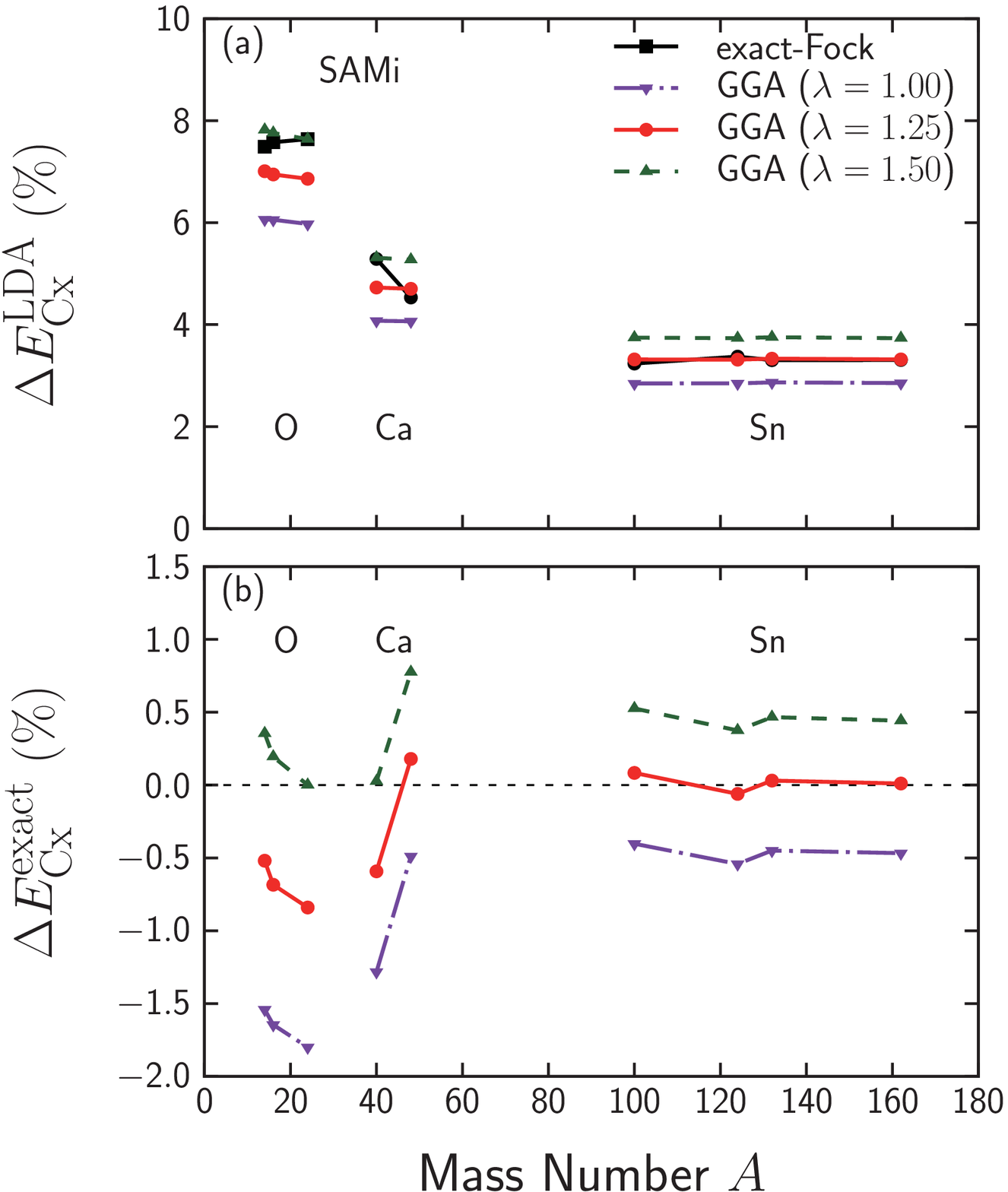}
  \caption{
    The same as Fig.~\ref{fig:systematic_dmagic} but for the $ \mathrm{O} $, $ \mathrm{Ca} $, and $ \mathrm{Sn} $ isotopes.}
  \label{fig:systematic_isotope}
\end{figure}

By comparing the results of $ \nuc{Ca}{40}{} $ and $ \nuc{Ca}{48}{} $, one may establish whether the factor $ \lambda $ has an isospin dependence, i.e.,~dependence on $ \left( N - Z \right) / A $.
To see the difference between the LDA and the PBE-GGA results along isotopic chains, 
$ \Delta E_{\urm{Cx}}^{\urm{LDA}} $ and $ \Delta E_{\urm{Cx}}^{\urm{exact}} $ for the $ \mathrm{O} $, $ \mathrm{Ca} $, and $ \mathrm{Sn} $ isotopes are shown as a function of $ A $ in Figs.~\ref{fig:systematic_isotope}(a) and \ref{fig:systematic_isotope}(b), respectively.
As shown in Fig.~\ref{fig:systematic_isotope}, the PBE-GGA with $ \lambda = 1.50 $ reproduces the exact-Fock results for all $ \mathrm{O} $ isotopes, and the PBE-GGA with $ \lambda = 1.25 $ reproduces the exact-Fock results for all $ \mathrm{Sn} $ isotopes.
For $ \mathrm{Ca} $ isotopes, $ \lambda = 1.50 $ works well in $ \nuc{Ca}{40}{} $, while $ \lambda = 1.25 $ works well in $ \nuc{Ca}{48}{} $.
Therefore, we cannot draw a firm conclusion on $ \mathrm{Ca} $ isotopes and on the isospin dependence of $ \lambda $.
We should remind that other open questions exist for $ \mathrm{Ca} $ isotopes:
for instance, the charge radius of $ \nuc{Ca}{48}{} $ is smaller than that of $ \nuc{Ca}{40}{} $ \cite{Angeli2013At.DataNucl.DataTables99_69},
whereas a nucleus with the larger mass number has a larger charge radius in most isotopic chains.
\par
Furthermore, in light nuclei, many properties are more sensitive to the shell structure, 
and thus even $ \lambda $ may be more sensitive to $ A $ and $ Z $.
In contrast, in medium-heavy and heavy nuclei the sensitivity to the shell structure is less pronounced. 
As a conclusion, $ \lambda $ does not seem to have an obvious isospin 
dependence, and $ \lambda = 1.25 $ reproduces the exact-Fock calculation well in a wide region of the nuclear chart.
\par
It should be noted that the PBE-GGA Coulomb potential is a local potential, and thus the numerical cost of the self-consistent calculation is $ O \left( N^3 \right) $,
while the exact-Fock Coulomb potential is a nonlocal potential and thus the numerical cost is $ O \left( N^4 \right) $.
Hence, the self-consistent calculations using the PBE-GGA functional with $ \lambda = 1.25 $ have a lower numerical cost and almost the same accuracy as the exact-Fock calculation.
\par
Before ending this section, let us stress a difference between electronic systems and nuclei.
In DFT, ideally, the total Coulomb energy, i.e.,~the sum of the Coulomb direct, exchange, and correlation energies, has a physical meaning while the same cannot be said of each separate contribution to the energy.
However, since the contribution of Coulomb correlations to EDFs are not considered, usually, in nuclear physics,
the Coulomb exchange term of the EDF discussed here should be required to reproduce the exact-Fock energy.
In contrast, in electron systems, the Coulomb exchange part of the EDF is not supposed to reproduce the exact-Fock energy.
Instead, the exchange and correlation terms of EDFs, together, ought to reproduce the total energy.
Therefore, the roles of the exchange term of EDFs are slightly different in the two cases.
Accordingly, the value of $ \lambda $ is different from one, yet it is expected to be of the same order.
\begin{table}[t]
  \centering
  \caption{Coulomb exchange energies $ E_{\urm{Cx}} $ for the doubly magic and semimagic nuclei calculated with the LDA and PBE-GGA Coulomb exchange functionals are compared with the exact-Fock energies. Units are $ \mathrm{MeV} $.
    For the nuclear part of EDF, the SAMi functional \cite{Roca-Maza2012Phys.Rev.C86_031306} is used.}
  \label{tab:sys_Ecx_dmagic}
  \begin{tabular}{lddddd}
    \hline \hline
    Nucleus & \multicolumn{1}{c}{LDA} & \multicolumn{1}{c}{Exact-Fock} & \multicolumn{1}{c}{PBE-GGA ($ \lambda = 1.00 $)} & \multicolumn{1}{c}{PBE-GGA ($ \lambda = 1.25 $)} & \multicolumn{1}{c}{PBE-GGA ($ \lambda = 1.50 $)} \\ \hline
    $ \nuc{He}{4}{} $   &  -0.627 &  -0.732 &  -0.701 &  -0.712 &  -0.722 \\
    $ \nuc{O}{14}{} $   &  -2.866 &  -3.098 &  -3.051 &  -3.082 &  -3.109 \\
    $ \nuc{O}{16}{} $   &  -2.854 &  -3.088 &  -3.038 &  -3.067 &  -3.094 \\
    $ \nuc{O}{24}{} $   &  -2.770 &  -2.999 &  -2.946 &  -2.974 &  -2.999 \\
    $ \nuc{Ca}{40}{} $  &  -7.558 &  -7.980 &  -7.879 &  -7.933 &  -7.982 \\
    $ \nuc{Ca}{48}{} $  &  -7.458 &  -7.812 &  -7.774 &  -7.826 &  -7.873 \\
    $ \nuc{Sn}{100}{} $ & -19.768 & -20.429 & -20.347 & -20.446 & -20.537 \\
    $ \nuc{Sn}{124}{} $ & -19.001 & -19.664 & -19.558 & -19.652 & -19.738 \\
    $ \nuc{Sn}{132}{} $ & -18.804 & -19.446 & -19.359 & -19.452 & -19.537 \\
    $ \nuc{Sn}{162}{} $ & -17.873 & -18.484 & -18.398 & -18.486 & -18.566 \\
    $ \nuc{Pb}{208}{} $ & -31.265 & -32.090 & -32.013 & -32.140 & -32.256 \\
    $ \nuc{126}{310}{}$ & -48.304 & -49.305 & -49.266 & -49.432 & -49.585 \\
    \hline \hline
  \end{tabular}
\end{table}
\subsection{Detailed analysis for $ \nuc{Pb}{208}{} $}
\label{subsec:detail}
\par
To understand in more detail, the ground-state properties of $ \nuc{Pb}{208}{} $ influenced by the Coulomb interaction will be discussed.
In this section, $ \lambda = 1.25 $ is used for the PBE-GGA Coulomb exchange functional.
\par
\begin{figure}[t]
  \centering
  \includegraphics[width=8cm]{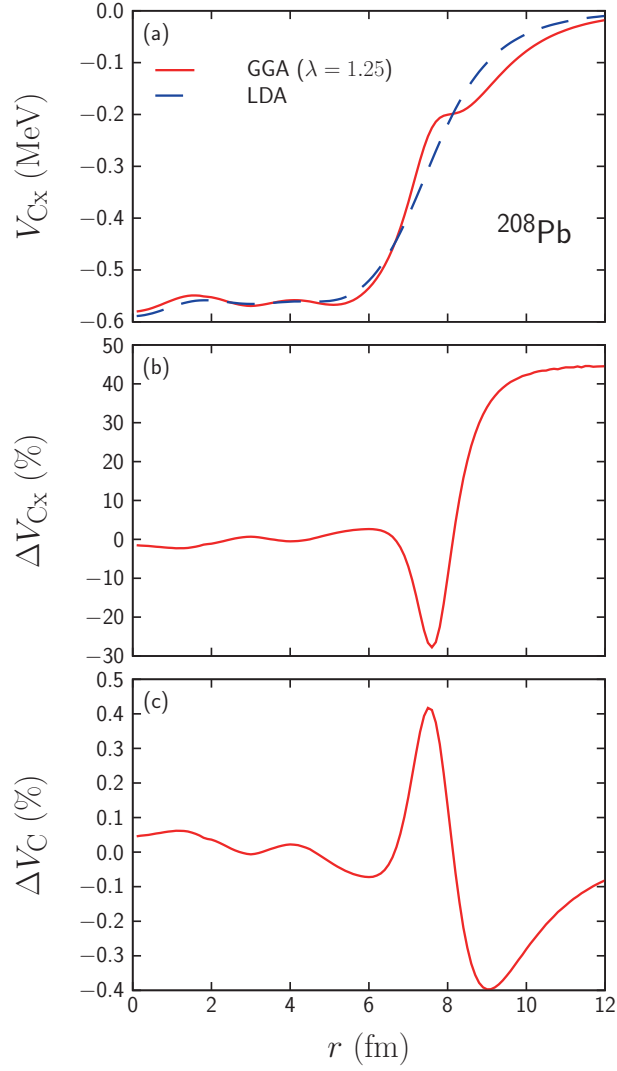}
  \caption{
    (a) The LDA (long-dashed line) and PBE-GGA (solid line) Coulomb exchange potentials $ V_{\urm{Cx}} $ as a function of $ r $.
    (b) The deviation of the PBE-GGA and LDA Coulomb exchange potential $ \Delta V_{\urm{Cx}} $ as a function of $ r $.
    (c) The same as panel (b) but for the total Coulomb potential $ \Delta V_{\urm{C}} $.}
  \label{fig:pb208_Coulomb}
\end{figure}

The LDA and PBE-GGA Coulomb exchange potentials $ V_{\urm{Cx}} $, after convergence, are shown as a function of $ r $ by means of long-dashed and solid lines, respectively, in Fig.~\ref{fig:pb208_Coulomb}(a).
To see clearly the difference between the results calculated by the PBE-GGA Coulomb exchange functional and those 
calculated within LDA, 
the deviation of the PBE-GGA and LDA Coulomb exchange potential, namely
\begin{equation}
  \label{eq:pot_diff}
  \Delta V_{\urm{Cx}}
  =
  \frac{V_{\urm{Cx}}^{\urm{GGA}} - V_{\urm{Cx}}^{\urm{LDA}}}{V_{\urm{Cx}}^{\urm{GGA}}}, 
\end{equation}
and the same relative deviation for the total Coulomb potential, $ \Delta V_{\urm{C}} $, are shown in Figs.~\ref{fig:pb208_Coulomb}(b) and \ref{fig:pb208_Coulomb}(c), respectively.
It is seen that the deviation between the PBE-GGA and LDA Coulomb exchange energy is significant;
in particular, it becomes $ -30 \, \% $ in the surface region and $ 40 \, \% $ in the tail region.
The dip in the surface region is because of the derivative of $ \rho_p $,
while the asymptotic behavior in the tail region is due to the saturation of the enhancement factor $ F $ as $ s $ increases.
However, the Coulomb exchange potential is quite weak compared with the Coulomb direct potential.
Thus, although the deviation is non-negligible for Coulomb exchange potential, 
that for the total Coulomb potential is less than $ 0.5 \, \% $.
\par
\begin{figure}[t]
  \centering
  \includegraphics[width=8cm]{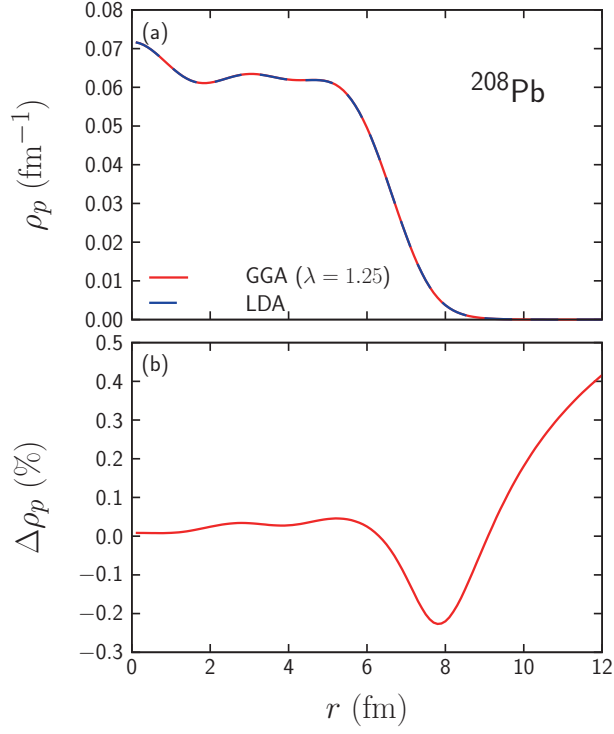}
  \caption{
    (a) Proton density distribution $ \rho_p $ for $ \nuc{Pb}{208}{} $ as a function of $ r $, where the results calculated from the LDA and PBE-GGA are shown by means of long-dashed and solid lines, respectively.
    (b) Deviation of the density from PBE-GGA with respect to that from LDA, $ \Delta \rho_p $, as a function of $ r $.}
  \label{fig:pb208_dens_proton}
\end{figure}
\begin{figure}[t]
  \centering
  \includegraphics[width=8cm]{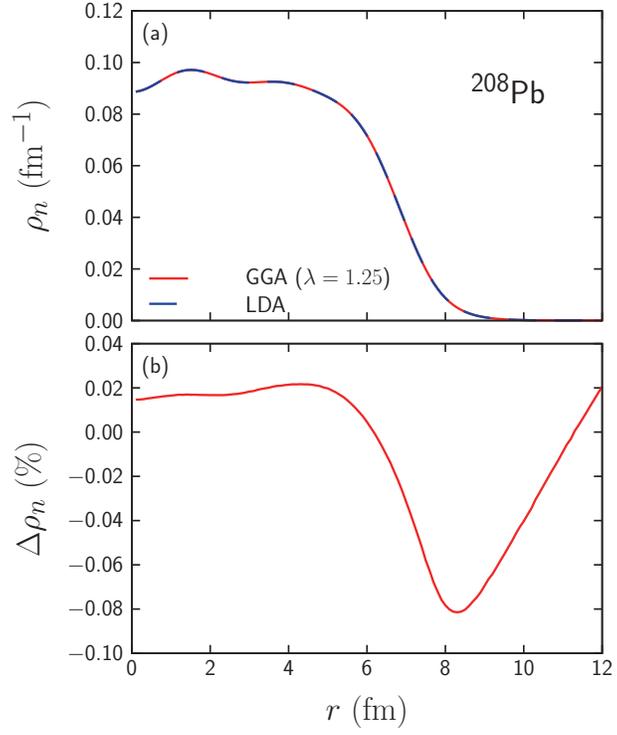}
  \caption{
    The same as Fig.~\ref{fig:pb208_dens_proton} but for neutron density distribution $ \rho_n $ 
and the corresponding deviation $ \Delta \rho_n $.}
  \label{fig:pb208_dens_neutron}
\end{figure}

The deviation between the density calculated using the PBE-GGA Coulomb exchange functional and the LDA functional, 
\begin{equation}
  \label{eq:dens_diff}
  \Delta \rho_{\tau}
  =
  \frac{\rho_{\tau}^{\urm{GGA}} - \rho_{\tau}^{\urm{LDA}}}{\rho_{\tau}^{\urm{GGA}}}
  \qquad
  \text{($ \tau = p , \, n $)},
\end{equation}
is shown as a function of $ r $ in Figs.~\ref{fig:pb208_dens_proton} and \ref{fig:pb208_dens_neutron}, respectively.
The proton densities calculated by employing the LDA and GGA approaches are identical within $ 0.5 \, \% $, and the neutron density within $ 0.1 \, \% $.
Since the Coulomb potential $ V_{\urm{C}} $ affects the proton single-particle orbitals and thus the density $ \rho_p $ directly,
the absolute value of $ \Delta \rho_p $ is of the same order as $ \Delta V_{\urm{C}} $,
and thus the $ 0.5 \, \% $ accuracy corresponds to that on $ \Delta V_{\urm{C}} $.
In contrast, since the Coulomb functional does not affect the neutron density directly, 
neutrons are affected through the change of the proton distribution caused by $ \Delta V_{\urm{C}} $ indirectly,
and as a result, the absolute value of $ \Delta \rho_n $ is one order of magnitude smaller than $ \Delta \rho_p $.
Even though they are tiny, both $ \Delta \rho_p $ and $ \Delta \rho_n $ are negative in the surface region, as the PBE-GGA Coulomb potential is larger than the LDA one there.
Moreover, since the LDA and PBE-GGA Coulomb exchange functionals give very similar proton densities $ \rho_p $,
the difference of the two Coulomb exchange potential $ \Delta V_{\urm{Cx}} $ comes from the enhancement factor 
rather than from the difference of the densities.
\par
The single-particle energies $ \epsilon_j $ for protons in $ \nuc{Pb}{208}{} $ calculated using either the LDA or the PBE-GGA Coulomb exchange functionals are shown in Table \ref{tab:pb208_sp_energy}.
Those from the exact-Fock term \cite{Roca-Maza2016Phys.Rev.C94_044313} are also shown.
Since the Coulomb potential changes quite a little as shown in Fig.~\ref{fig:pb208_Coulomb}(c), the single-particle energies also change quite a little.
The differences in $ \epsilon_j $ calculated either within LDA or PBE-GGA are less than $ 10 \, \mathrm{keV} $,
while the differences between those calculated either within LDA or by using exact-Fock term are more than $ 100 \, \mathrm{keV} $.
Even though the PBE-GGA Coulomb exchange functional does not change the single-particle energies $ \epsilon_j $, the exchange Coulomb energy $ E_{\urm{Cx}} $ in the PBE-GGA is almost the same as the exact-Fock energy.
To understand the reason, the exchange Coulomb energy $ E_{\urm{Cx}} $, the total energy per particle $ E_{\urm{tot}} /A $, the sum of the single-particle energies and kinetic energies per particle $ \sum_j \left( \epsilon_j + \tau_j \right) /2A $, and the rearrangement term of the total energy per particle $ E_{\urm{rea}} /A $  for $ \nuc{Pb}{208}{} $ are shown in Table \ref{tab:pb208_total_energy}.
The total energy is calculated by means of Eq.~\eqref{eq:energy_gs2}.
It is seen that both the total energy and the Coulomb exchange energy are almost the same when comparing the exact-Fock
and the PBE-GGA
(differences are $ 0.001 $ and $ 0.05 \, \mathrm{MeV} $, respectively), and they differ more with respect to LDA.
However,  
$ \sum_j \left( \epsilon_j + \tau_j \right) /2A $ and $ E_{\urm{rea}} /A $ 
are more similar when comparing PBE-GGA and LDA than when comparing PBE-GGA and LDA.
In other words, nontrivial cancellations are at work.
\begin{table}[t]
  \centering
  \caption{
    Single-particle energies for protons in $ \nuc{Pb}{208}{} $ calculated from the LDA and PBE-GGA Coulomb exchange functionals.
    Those from the exact-Fock calculation \cite{Roca-Maza2016Phys.Rev.C94_044313} are also shown.
    Units are $ \mathrm{MeV} $.}
  \label{tab:pb208_sp_energy}
  \begin{tabular}{lddd}
    \hline \hline
    Orbital & \multicolumn{1}{c}{Exact-Fock \cite{Roca-Maza2016Phys.Rev.C94_044313}} & \multicolumn{1}{c}{LDA} & \multicolumn{1}{c}{GGA ($ \lambda = 1.25 $)} \\ \hline
    $ 1s_{1/2} $ & -45.501 & -44.980 & -44.983 \\
    $ 1p_{3/2} $ & -39.863 & -39.387 & -39.390 \\
    $ 1p_{1/2} $ & -39.574 & -39.107 & -39.111 \\
    $ 1d_{5/2} $ & -32.903 & -32.482 & -32.485 \\
    $ 1d_{3/2} $ & -32.209 & -31.815 & -31.817 \\
    $ 2s_{1/2} $ & -28.899 & -28.509 & -28.507 \\
    $ 1f_{7/2} $ & -25.045 & -24.692 & -24.693 \\
    $ 1f_{5/2} $ & -23.648 & -23.353 & -23.353 \\
    $ 2p_{3/2} $ & -19.702 & -19.411 & -19.406 \\
    $ 2p_{1/2} $ & -18.906 & -18.626 & -18.621 \\
    $ 1g_{9/2} $ & -16.605 & -16.338 & -16.336 \\
    $ 1g_{7/2} $ & -14.175 & -14.019 & -14.017 \\
    $ 2d_{5/2} $ & -10.411 & -10.255 & -10.246 \\
    $ 2d_{3/2} $ &  -8.897 &  -8.846 &  -8.837 \\
    $ 3s_{1/2} $ &  -7.813 &  -7.673 &  -7.660 \\
    $ 1h_{11/2}$ &  -7.802 &  -7.663 &  -7.658 \\
    \hline \hline
  \end{tabular}
\end{table}
\begin{table}[t]
  \centering
  \caption{
    The exchange Coulomb energy $ E_{\urm{Cx}} $, the total energy per particle $ E_{\urm{tot}} /A $, the sum of the single-particle energies and kinetic energies per particle $ \sum_j \left( \epsilon_j + \tau_j \right) /2A $, and the rearrangement term of the energy per particle $ E_{\urm{rea}} /A $. All values correspond to $ \nuc{Pb}{208}{} $ and 
    units are $ \mathrm{MeV} $.}
  \label{tab:pb208_total_energy}
  \begin{tabular}{lddd}
    \hline \hline
    & \multicolumn{1}{c}{Exact-Fock} & \multicolumn{1}{c}{LDA} & \multicolumn{1}{c}{GGA ($ \lambda = 1.25 $)} \\ \hline
    $ E_{\urm{Cx}} $ & -32.090 & -31.265 & -32.140 \\ \hline
    $ E_{\urm{tot}} / A $ & -7.872 & -7.868 & -7.873 \\
    $ \sum_j \left( \epsilon_j + \tau_j \right) / 2A $ & -2.225 & -2.171 & -2.169 \\
    $ E_{\urm{rea}} / A $ & -5.647 & -5.697 & -5.703 \\ \hline
    \hline
  \end{tabular}
\end{table}
\clearpage
%
\section{Conclusions and Perspectives}
\label{sec:conc}
\par
In this paper, we have applied the PBE-GGA Coulomb exchange functional to the self-consistent Skyrme Hartree-Fock calculations for atomic nuclei.
To reproduce the exact-Fock energy, one of the PBE-GGA parameters $ \mu $ is changed to $ \lambda \mu $.
It is found that $ \lambda = 1.25 $ is the most suitable value for nuclei in a wide region of the nuclear chart; $ \lambda $ does not have an obvious isospin dependence,
although there are some open questions for the behavior of the $ \mathrm{Ca} $ isotopes and for the suitable values for the light- or superheavy nuclei.
It should be emphasized that the numerical cost of the self-consistent calculations with the PBE-GGA exchange functional is $ O \left( N^3 \right) $, whereas that with the exact-Fock term is $ O \left( N^4 \right) $.
\par
For the Coulomb exchange energy, it is found that the deviation between the PBE-GGA and the LDA results, $ \Delta E_{\urm{Cx}}^{\urm{LDA}} $, ranges from around $ 12 \, \% $ in $ \nuc{He}{4}{} $ to $ 2 \, \% $ in $ \nuc{126}{310}{} $.
This behavior is similar to $ \Delta E_{\urm{Cx}}^{\urm{LDA}} $ for the exact-Fock.
Compared with the exact-Fock calculation, the deviation between the PBE-GGA and the exact-Fock results, $ \Delta E_{\urm{Cx}}^{\urm{exact}} $,
ranges from $ -3 \, \% $ in $ \nuc{He}{4}{} $ to less than $ 1 \, \% $ in $ \nuc{126}{310}{} $,
which means the PBE-GGA with $ \lambda = 1.25 $ reproduces the exact-Fock energy within $ \approx 100 \, \mathrm{keV} $ accuracy.
In contrast, it is found that the proton and neutron density distributions and the single-particle energies calculated within PBE-GGA give almost the same results as those calculated within LDA
since the Coulomb potential changes quite a little from LDA to PBE-GGA.
It has been shown that the Coulomb exchange energy in the PBE-GGA reproduces the exact-Fock energy due to the effect of the rearrangement term $ E_{\urm{rea}} $, 
not only for $ \nuc{Pb}{208}{} $ but also for the other nuclei, although we have not shown the details explicitly.
\par
It is known that the error of the total Coulomb energy can be separable into two parts:
the density-driven error and functional-driven error \cite{Kim2013Phys.Rev.Lett.111_073003}.
The latter comes from the difference between the ``exact'' functional $ E \left[ \rho_{\urm{gs}} \right] $ and the ``approximated'' functional $ \tilde{E} \left[ \rho_{\urm{gs}} \right] $,
and the former is the remaining part and comes from the difference between the exact ground-state density $ \rho_{\urm{gs}} $ and the calculated ground-state density $ \tilde{\rho}_{\urm{gs}} $.
For the Coulomb exchange functional, between the LDA and the PBE-GGA, the calculated ground-state densities are almost the same,  
and the exact-Fock energy can be calculated by means of first-order perturbation theory.
Thus, the difference between the PBE-GGA energy and the exact-Fock one dominantly comes from the functional-driven error.
\par
The finite-size effect of protons is an interesting topic.
Since electrons are elementary particles and do not have a finite radius,
charge distributions and electron distributions are identical to each other in electron systems.
In contrast, since protons have a finite radius \cite{Mohr2016Rev.Mod.Phys.88_035009},
charge distributions $ \rho_{\urm{ch}} $ and proton distributions $ \rho_p $ differ from each other.
Even though the self-consistent calculations of the DFT or Hartree-Fock type for nuclear structure usually do not consider this difference,
it is known that the finite-size effect of protons is not negligible, e.g.,~in the study of the energy of the Isobaric Analog State \cite{Roca-Maza2018Phys.Rev.Lett.120_202501}.
One more essential point is that it is difficult to consider the form factor of protons for the finite-size effect in the exact-Fock term as the single-particle wave functions are used.
In contrast, one is able to consider this effect in the PBE-GGA Coulomb exchange functional since the functional is written only in terms of the density.
\par
A more challenging topic for future work is the Coulomb correlation part of EDFs for nuclear systems.
In electron systems, Coulomb correlations have been discussed for decades.
However, these functionals are not applicable to nuclear systems \cite{Naito2018Phys.Rev.C97_044319}.
This is because the correlation energy in nuclear systems is mainly caused by the attractive nuclear interaction,
whereas that in electron systems is mainly caused by the repulsive Coulomb interaction,
and thus, the Coulomb correlation energy in nuclear systems have the opposite sign with respect to the electron systems \cite{Naito2018Phys.Rev.C97_044319,Bulgac1996Nucl.Phys.A601_103,Bulgac1999Phys.Lett.B469_1}.
Therefore, to investigate the Coulomb correlation EDFs for nuclear systems is a very important topic for nuclear systems in the future. 
\begin{acknowledgments}
\par
The authors appreciate Ryosuke Akashi, Shinji Tsuneyuki, and Enrico Vigezzi for stimulating discussions and valuable comments.
T.N.~and H.L.~would like to thank the RIKEN iTHEMS program
and the JSPS-NSFC Bilateral Program for Joint Research Project on Nuclear mass and life for unravelling mysteries of the $ r $-process.
T.~N.~acknowledges the financial support from Computational Science Alliance, The University of Tokyo.
H.L.~acknowledges the JSPS Grant-in-Aid for Early-Career Scientists under Grant No.~18K13549.
G.C.~and X.R.-M.~acknowledge funding from the European Union's Horizon 2020 research and innovation program under Grant No.~654002.
\end{acknowledgments}
%
\appendix
\section{Potential and Rearrangement Term in Generalized Gradient Approximation}
\label{sec:app}
\par
\begin{figure}[t]
  \centering
  \includegraphics[width=8cm]{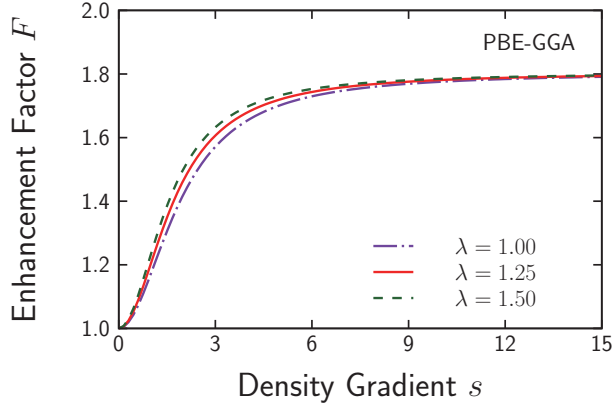}
  \caption{
    Enhancement factors $ F $ for the PBE-GGA \cite{Perdew1996Phys.Rev.Lett.77_3865} for $ \lambda = 1.00 $, $ 1.25 $, and $ 1.50 $ as functions of $ s $ are shown in dot-dashed, solid, and dashed lines, respectively.}
  \label{fig:enhancement}
\end{figure}

In the self-consistent calculation, a potential form of the Coulomb exchange functional is required.
Since the LDA Coulomb exchange functional is
\begin{equation}
  \label{eq:LDA_E}
  E_{\urm{Cx}}^{\urm{LDA}}
  \left[
    \rho
  \right]
  =
  -
  \frac{3}{4}
  \frac{e^2}{4 \pi \epsilon_0}
  \left( \frac{3}{\pi} \right)^{1/3}
  \int
  \left[
    \rho \left( \ve{r} \right)
  \right]^{4/3}
  \, d \ve{r},
\end{equation}
the LDA Coulomb exchange potential is
\begin{align}
  V_{\urm{Cx}}^{\urm{LDA}} \left( \ve{r} \right)
  & =
    \frac{\delta E_{\urm{Cx}}^{\urm{LDA}} \left[ \rho \left( \ve{r} \right) \right]}{\delta \rho \left( \ve{r} \right)} \notag \\
  & =
    \epsilon_{\urm{Cx}}^{\urm{LDA}} \left( \rho \right) 
    +
    \rho \left( \ve{r} \right)
    \frac{\partial \epsilon_{\urm{Cx}}^{\urm{LDA}} \left( \rho \right)}{\partial \rho} \notag \\
  & =
    -
    \frac{e^2}{4 \pi \epsilon_0}
    \left( \frac{3}{\pi} \right)^{1/3}
    \left[
    \rho \left( \ve{r} \right)
    \right]^{1/3},
\end{align}
where $ \epsilon_{\urm{Cx}}^{\urm{LDA}} $ is the exchange energy density 
\begin{equation}
  \epsilon_{\urm{Cx}}^{\urm{LDA}} \left( \rho \right)
  =
  -
  \frac{3}{4}
  \frac{e^2}{4 \pi \epsilon_0}
  \left( \frac{3}{\pi} \right)^{1/3}
  \rho^{1/3}.
\end{equation}
To go beyond the LDA, the GGA Coulomb exchange functional and energy density are
\begin{align}
  E_{\urm{Cx}}^{\urm{GGA}}
  \left[
  \rho
  \right]
  & =
    -
    \frac{3}{4}
    \frac{e^2}{4 \pi \epsilon_0}
    \left( \frac{3}{\pi} \right)^{1/3}
    \int
    \left[
    \rho \left( \ve{r} \right)
    \right]^{4/3}
    F \left( s \left( \ve{r} \right) \right)
    \, d \ve{r},
    \label{eq:GGA_E} \\
  \epsilon_{\urm{Cx}}^{\urm{GGA}} \left( \rho \right)
  & =
    -
    \frac{3}{4}
    \frac{e^2}{4 \pi \epsilon_0}
    \left( \frac{3}{\pi} \right)^{1/3}
    \rho^{1/3}
    F \left( s \right),
\end{align}
respectively.
Thus, the GGA Coulomb exchange potential is \cite{Carmona-Espindola2015J.Chem.Phys.142_054105}
\begin{align}
  V_{\urm{Cx}}^{\urm{GGA}} \left( \ve{r} \right)
  = & \,
      \frac{\delta E_{\urm{Cx}}^{\urm{GGA}} \left[ \rho \left( \ve{r} \right) \right]}{\delta \rho \left( \ve{r} \right)} \notag \\
  = & \,
      \epsilon_{\urm{Cx}}^{\urm{GGA}} \left( \rho \right)
      +
      \rho
      \frac{\partial \epsilon_{\urm{Cx}}^{\urm{GGA}} \left( \rho \right)}{\partial \rho} 
      -
      \Nabla
      \cdot
      \left(
      \rho
      \frac{\partial \epsilon_{\urm{Cx}}^{\urm{GGA}} \left( \rho \right)}{\partial \Nabla \rho} 
      \right) \notag \\
  = & \,
      V_{\urm{Cx}}^{\urm{LDA}} \left( \ve{r} \right) \,
      F \left( s \right) \notag \\
    & +
      \frac{3}{4}
      V_{\urm{Cx}}^{\urm{LDA}} \left( \ve{r} \right) \,
      \left(
      \frac{\Nabla \rho \cdot \Nabla \left| \Nabla \rho \right|}
      {2 k_{\urm{F}} \left| \Nabla \rho \right|^2}
      -
      \frac{4}{3} s
      -
      \frac{1}{2 k_{\urm{F}}}
      \frac{\nabla^2 \rho}{\left| \Nabla \rho \right|}
      \right)
      \frac{dF \left(s \right)}{ds} \notag \\
    & - 
      \frac{3}{4}
      V_{\urm{Cx}}^{\urm{LDA}} \left( \ve{r} \right) \,
      \left(
      \frac{\Nabla \rho \cdot \Nabla \left| \Nabla \rho \right|}
      {\left( 2 k_{\urm{F}} \right)^2 \rho \left| \Nabla \rho \right|}
      -
      \frac{4}{3} s^2
      \right)
      \frac{d^2 F \left(s \right)}{ds^2},
      \label{eq:pot_GGA}
\end{align}
where the enhancement factors $ F $ for the PBE-GGA Coulomb exchange functional \cite{Perdew1996Phys.Rev.Lett.77_3865} for $ \lambda = 1.00 $, $ 1.25 $, and $ 1.50 $ as functions of $ s $ are shown in dot-dashed, solid, and dashed lines, respectively, in Fig.~\ref{fig:enhancement}.
The GGA Coulomb exchange potential shown in Eq.~\eqref{eq:pot_GGA} is applicable for general GGA Coulomb exchange functionals, including the PBE-GGA Coulomb exchange functionals.
Under the assumption of spherical symmetry, Eq.~\eqref{eq:pot_GGA} is simplified as 
\begin{equation}
  V_{\urm{Cx}}^{\urm{GGA}} \left( r \right)
  =
  V_{\urm{Cx}}^{\urm{LDA}} \left( r \right)
  \left[
    F \left( s \right)
    -
    \left(
      s
      +
      \frac{3}{4}
      \frac{1}{k_{\urm{F}} r}
    \right)
    \frac{dF}{ds}
    +
    \left(
      s^2
      -
      \frac{3}{4}
      \frac{\rho''}{4 \rho k_{\urm{F}}^2}
    \right)
    \frac{d^2F}{ds^2}
  \right],
\end{equation}
where $ \rho'' = d^2 \rho \left( r \right) / dr^2 $.
\par
Next, the total energy is considered.
On the one hand, in the original DFT, the total energy is written as \cite{Engel2011_Springer-Verlag}
\begin{equation}
  \label{eq:gs_dft}
  E_{\urm{gs}}
  =
  \sum_j
  \epsilon_j 
  +
  E_{\urm{xc}} \left[ \rho_{\urm{gs}} \right]
  -
  \int
  V_{\urm{xc}} \left( \ve{r} \right) \,
  \rho_{\urm{gs}} \left( \ve{r} \right) \,
  d \ve{r}
  +
  \frac{1}{2}
  \iint
  V_{\urm{int}} \left( \ve{r}, \ve{r}' \right) \,
  \rho_{\urm{gs}} \left( \ve{r} \right) \,
  \rho_{\urm{gs}} \left( \ve{r}' \right) \,
  d \ve{r} \, d \ve{r}',
\end{equation}
where
$ V_{\urm{int}} $ is the two-body interaction,
$ E_{\urm{xc}} $ is the exchange-correlation functional,
which shows the exchange energy and the remaining part of the total energy,
and $ V_{\urm{xc}} $ is the exchange-correlation potential defined as 
$ V_{\urm{xc}} = \left. \delta E_{\urm{xc}} / \delta \rho \right|_{\rho = \rho_{\uurm{gs}}} $.
On the other hand, in the Skyrme Hartree-Fock calculation, the total energy reads
\begin{equation}
  \label{eq:gs_sky}
  E_{\urm{gs}}
  =
  \frac{1}{2}
  \sum_j
  \left(
    \epsilon_j + \tau_j
  \right)
  +
  E_{\urm{rea}},
\end{equation}
where $ \epsilon_j $ and $ \tau_j $ are the single-particle energy and the single-particle kinetic energy, respectively, and $ E_{\urm{rea}} $ is the rearrangement term \cite{Vautherin1972Phys.Rev.C5_626}.
The two expressions for the total energy given by Eqs.~\eqref{eq:gs_dft} and \eqref{eq:gs_sky} should be identical to each other.
Since the single-particle Kohn-Sham Hamiltonian is
\begin{equation}
  \hat{h}
  =
  \hat{t}
  +
  V_{\urm{xc}} \left( \ve{r} \right)
  + 
  \int
  V_{\urm{int}} \left( \ve{r}, \ve{r}' \right) \,
  \rho_{\urm{gs}} \left( \ve{r}' \right) \,
  d \ve{r}',
\end{equation}
where $ \hat{t} $ is the single-particle kinetic operator,
the equation
\begin{equation}
  \sum_j
  \left(
    \epsilon_j - \tau_j
  \right)
  =
  \int
  V_{\urm{xc}} \left( \ve{r} \right) \,
  \rho_{\urm{gs}} \left( \ve{r} \right) \,
  d \ve{r}
  +
  \iint
  V_{\urm{int}} \left( \ve{r}, \ve{r}' \right) \,
  \rho_{\urm{gs}} \left( \ve{r} \right) \,
  \rho_{\urm{gs}} \left( \ve{r}' \right) \,
  d \ve{r} \, d \ve{r}'
\end{equation}
follows,
and therefore Eq.~\eqref{eq:gs_dft} reads
\begin{equation}
  E_{\urm{gs}}
  =
  \frac{1}{2}
  \sum_j
  \left(
    \epsilon_j
    +
    \tau_j
  \right)
  +
  E_{\urm{xc}} \left[ \rho_{\urm{gs}} \right]
  -
  \frac{1}{2}
  \int
  V_{\urm{xc}} \left( \ve{r} \right) \,
  \rho_{\urm{gs}} \left( \ve{r} \right) \,
  d \ve{r}.
\end{equation}
As compared with Eq.~\eqref{eq:gs_sky}, the rearrangement term is
\begin{equation}
  E_{\urm{rea}}
  =
  E_{\urm{xc}} \left[ \rho_{\urm{gs}} \right]
  -
  \frac{1}{2}
  \int
  V_{\urm{xc}} \left( \ve{r} \right) \,
  \rho_{\urm{gs}} \left( \ve{r} \right) \,
  d \ve{r}.
\end{equation}
%
\bibliography{0_proton_scf}

\begin{thebibliography}{42}%
\makeatletter
\providecommand \@ifxundefined [1]{%
 \@ifx{#1\undefined}
}%
\providecommand \@ifnum [1]{%
 \ifnum #1\expandafter \@firstoftwo
 \else \expandafter \@secondoftwo
 \fi
}%
\providecommand \@ifx [1]{%
 \ifx #1\expandafter \@firstoftwo
 \else \expandafter \@secondoftwo
 \fi
}%
\providecommand \natexlab [1]{#1}%
\providecommand \enquote  [1]{``#1''}%
\providecommand \bibnamefont  [1]{#1}%
\providecommand \bibfnamefont [1]{#1}%
\providecommand \citenamefont [1]{#1}%
\providecommand \href@noop [0]{\@secondoftwo}%
\providecommand \href [0]{\begingroup \@sanitize@url \@href}%
\providecommand \@href[1]{\@@startlink{#1}\@@href}%
\providecommand \@@href[1]{\endgroup#1\@@endlink}%
\providecommand \@sanitize@url [0]{\catcode `\\12\catcode `\$12\catcode
  `\&12\catcode `\#12\catcode `\^12\catcode `\_12\catcode `\%12\relax}%
\providecommand \@@startlink[1]{}%
\providecommand \@@endlink[0]{}%
\providecommand \url  [0]{\begingroup\@sanitize@url \@url }%
\providecommand \@url [1]{\endgroup\@href {#1}{\urlprefix }}%
\providecommand \urlprefix  [0]{URL }%
\providecommand \Eprint [0]{\href }%
\providecommand \doibase [0]{http://dx.doi.org/}%
\providecommand \selectlanguage [0]{\@gobble}%
\providecommand \bibinfo  [0]{\@secondoftwo}%
\providecommand \bibfield  [0]{\@secondoftwo}%
\providecommand \translation [1]{[#1]}%
\providecommand \BibitemOpen [0]{}%
\providecommand \bibitemStop [0]{}%
\providecommand \bibitemNoStop [0]{.\EOS\space}%
\providecommand \EOS [0]{\spacefactor3000\relax}%
\providecommand \BibitemShut  [1]{\csname bibitem#1\endcsname}%
\let\auto@bib@innerbib\@empty
\bibitem [{\citenamefont {Nolen}\ and\ \citenamefont
  {Schiffer}(1969)}]{Nolen1969Annu.Rev.Nucl.Sci.19_471}%
  \BibitemOpen
  \bibfield  {author} {\bibinfo {author} {\bibfnamefont {J.~A.}\ \bibnamefont
  {Nolen}, \bibfnamefont {Jr.}}\ and\ \bibinfo {author} {\bibfnamefont {J.~P.}\
  \bibnamefont {Schiffer}},\ }\href {\doibase
  10.1146/annurev.ns.19.120169.002351} {\bibfield  {journal} {\bibinfo
  {journal} {Annu. Rev. Nucl. Sci.}\ }\textbf {\bibinfo {volume} {19}},\
  \bibinfo {pages} {471} (\bibinfo {year} {1969})}\BibitemShut {NoStop}%
\bibitem [{\citenamefont {J\"{a}necke}(1965)}]{Jaenecke1965Nucl.Phys.73_97}%
  \BibitemOpen
  \bibfield  {author} {\bibinfo {author} {\bibfnamefont {J.}~\bibnamefont
  {J\"{a}necke}},\ }\href {\doibase 10.1016/0029-5582(65)90157-4} {\bibfield
  {journal} {\bibinfo  {journal} {Nucl. Phys.}\ }\textbf {\bibinfo {volume}
  {73}},\ \bibinfo {pages} {97} (\bibinfo {year} {1965})}\BibitemShut {NoStop}%
\bibitem [{\citenamefont {Shlomo}(1978)}]{Shlomo1978Rep.Prog.Phys.41_957}%
  \BibitemOpen
  \bibfield  {author} {\bibinfo {author} {\bibfnamefont {S.}~\bibnamefont
  {Shlomo}},\ }\href {\doibase 10.1088/0034-4885/41/7/001} {\bibfield
  {journal} {\bibinfo  {journal} {Rep. Prog. Phys.}\ }\textbf {\bibinfo
  {volume} {41}},\ \bibinfo {pages} {957} (\bibinfo {year} {1978})}\BibitemShut
  {NoStop}%
\bibitem [{\citenamefont {Roca-Maza}\ \emph {et~al.}(2018)\citenamefont
  {Roca-Maza}, \citenamefont {Col\`{o}},\ and\ \citenamefont
  {Sagawa}}]{Roca-Maza2018Phys.Rev.Lett.120_202501}%
  \BibitemOpen
  \bibfield  {author} {\bibinfo {author} {\bibfnamefont {X.}~\bibnamefont
  {Roca-Maza}}, \bibinfo {author} {\bibfnamefont {G.}~\bibnamefont {Col\`{o}}},
  \ and\ \bibinfo {author} {\bibfnamefont {H.}~\bibnamefont {Sagawa}},\ }\href
  {\doibase 10.1103/PhysRevLett.120.202501} {\bibfield  {journal} {\bibinfo
  {journal} {Phys. Rev. Lett.}\ }\textbf {\bibinfo {volume} {120}},\ \bibinfo
  {pages} {202501} (\bibinfo {year} {2018})}\BibitemShut {NoStop}%
\bibitem [{\citenamefont {Coon}\ and\ \citenamefont
  {Scadron}(1982)}]{Coon1982Phys.Rev.C26_2402}%
  \BibitemOpen
  \bibfield  {author} {\bibinfo {author} {\bibfnamefont {S.~A.}\ \bibnamefont
  {Coon}}\ and\ \bibinfo {author} {\bibfnamefont {M.~D.}\ \bibnamefont
  {Scadron}},\ }\href {\doibase 10.1103/PhysRevC.26.2402} {\bibfield  {journal}
  {\bibinfo  {journal} {Phys. Rev. C}\ }\textbf {\bibinfo {volume} {26}},\
  \bibinfo {pages} {2402} (\bibinfo {year} {1982})}\BibitemShut {NoStop}%
\bibitem [{\citenamefont {Liang}\ \emph {et~al.}(2009)\citenamefont {Liang},
  \citenamefont {Van~Giai},\ and\ \citenamefont
  {Meng}}]{Liang2009Phys.Rev.C79_064316}%
  \BibitemOpen
  \bibfield  {author} {\bibinfo {author} {\bibfnamefont {H.}~\bibnamefont
  {Liang}}, \bibinfo {author} {\bibfnamefont {N.}~\bibnamefont {Van~Giai}}, \
  and\ \bibinfo {author} {\bibfnamefont {J.}~\bibnamefont {Meng}},\ }\href
  {\doibase 10.1103/PhysRevC.79.064316} {\bibfield  {journal} {\bibinfo
  {journal} {Phys. Rev. C}\ }\textbf {\bibinfo {volume} {79}},\ \bibinfo
  {pages} {064316} (\bibinfo {year} {2009})}\BibitemShut {NoStop}%
\bibitem [{\citenamefont {Hardy}\ and\ \citenamefont
  {Towner}(2015)}]{Hardy2015Phys.Rev.C91_025501}%
  \BibitemOpen
  \bibfield  {author} {\bibinfo {author} {\bibfnamefont {J.~C.}\ \bibnamefont
  {Hardy}}\ and\ \bibinfo {author} {\bibfnamefont {I.~S.}\ \bibnamefont
  {Towner}},\ }\href {\doibase 10.1103/PhysRevC.91.025501} {\bibfield
  {journal} {\bibinfo  {journal} {Phys. Rev. C}\ }\textbf {\bibinfo {volume}
  {91}},\ \bibinfo {pages} {025501} (\bibinfo {year} {2015})}\BibitemShut
  {NoStop}%
\bibitem [{\citenamefont {Titin-Schnaider}\ and\ \citenamefont
  {Quentin}(1974)}]{Titin-Schnaider1974Phys.Lett.B49_397}%
  \BibitemOpen
  \bibfield  {author} {\bibinfo {author} {\bibfnamefont {C.}~\bibnamefont
  {Titin-Schnaider}}\ and\ \bibinfo {author} {\bibfnamefont {P.}~\bibnamefont
  {Quentin}},\ }\href {\doibase 10.1016/0370-2693(74)90617-0} {\bibfield
  {journal} {\bibinfo  {journal} {Phys. Lett. B}\ }\textbf {\bibinfo {volume}
  {49}},\ \bibinfo {pages} {397} (\bibinfo {year} {1974})}\BibitemShut
  {NoStop}%
\bibitem [{\citenamefont {Anguiano}\ \emph {et~al.}(2001)\citenamefont
  {Anguiano}, \citenamefont {Egido},\ and\ \citenamefont
  {Robledo}}]{Anguiano2001Nucl.Phys.A683_227}%
  \BibitemOpen
  \bibfield  {author} {\bibinfo {author} {\bibfnamefont {M.}~\bibnamefont
  {Anguiano}}, \bibinfo {author} {\bibfnamefont {J.}~\bibnamefont {Egido}}, \
  and\ \bibinfo {author} {\bibfnamefont {L.}~\bibnamefont {Robledo}},\ }\href
  {\doibase 10.1016/S0375-9474(00)00445-0} {\bibfield  {journal} {\bibinfo
  {journal} {Nucl. Phys. A}\ }\textbf {\bibinfo {volume} {683}},\ \bibinfo
  {pages} {227} (\bibinfo {year} {2001})}\BibitemShut {NoStop}%
\bibitem [{\citenamefont {Skalski}(2001)}]{Skalski2001Phys.Rev.C63_024312}%
  \BibitemOpen
  \bibfield  {author} {\bibinfo {author} {\bibfnamefont {J.}~\bibnamefont
  {Skalski}},\ }\href {\doibase 10.1103/PhysRevC.63.024312} {\bibfield
  {journal} {\bibinfo  {journal} {Phys. Rev. C}\ }\textbf {\bibinfo {volume}
  {63}},\ \bibinfo {pages} {024312} (\bibinfo {year} {2001})}\BibitemShut
  {NoStop}%
\bibitem [{\citenamefont {Le~Bloas}\ \emph {et~al.}(2011)\citenamefont
  {Le~Bloas}, \citenamefont {Koh}, \citenamefont {Quentin}, \citenamefont
  {Bonneau},\ and\ \citenamefont {Ithnin}}]{LeBloas2011Phys.Rev.C84_014310}%
  \BibitemOpen
  \bibfield  {author} {\bibinfo {author} {\bibfnamefont {J.}~\bibnamefont
  {Le~Bloas}}, \bibinfo {author} {\bibfnamefont {M.-H.}\ \bibnamefont {Koh}},
  \bibinfo {author} {\bibfnamefont {P.}~\bibnamefont {Quentin}}, \bibinfo
  {author} {\bibfnamefont {L.}~\bibnamefont {Bonneau}}, \ and\ \bibinfo
  {author} {\bibfnamefont {J.~I.~A.}\ \bibnamefont {Ithnin}},\ }\href {\doibase
  10.1103/PhysRevC.84.014310} {\bibfield  {journal} {\bibinfo  {journal} {Phys.
  Rev. C}\ }\textbf {\bibinfo {volume} {84}},\ \bibinfo {pages} {014310}
  (\bibinfo {year} {2011})}\BibitemShut {NoStop}%
\bibitem [{\citenamefont {Roca-Maza}\ \emph {et~al.}(2016)\citenamefont
  {Roca-Maza}, \citenamefont {Cao}, \citenamefont {Col\`{o}},\ and\
  \citenamefont {Sagawa}}]{Roca-Maza2016Phys.Rev.C94_044313}%
  \BibitemOpen
  \bibfield  {author} {\bibinfo {author} {\bibfnamefont {X.}~\bibnamefont
  {Roca-Maza}}, \bibinfo {author} {\bibfnamefont {L.-G.}\ \bibnamefont {Cao}},
  \bibinfo {author} {\bibfnamefont {G.}~\bibnamefont {Col\`{o}}}, \ and\
  \bibinfo {author} {\bibfnamefont {H.}~\bibnamefont {Sagawa}},\ }\href
  {\doibase 10.1103/PhysRevC.94.044313} {\bibfield  {journal} {\bibinfo
  {journal} {Phys. Rev. C}\ }\textbf {\bibinfo {volume} {94}},\ \bibinfo
  {pages} {044313} (\bibinfo {year} {2016})}\BibitemShut {NoStop}%
\bibitem [{\citenamefont {Long}\ \emph {et~al.}(2006)\citenamefont {Long},
  \citenamefont {Van~Giai},\ and\ \citenamefont
  {Meng}}]{Long2006Phys.Lett.B640_150}%
  \BibitemOpen
  \bibfield  {author} {\bibinfo {author} {\bibfnamefont {W.-H.}\ \bibnamefont
  {Long}}, \bibinfo {author} {\bibfnamefont {N.}~\bibnamefont {Van~Giai}}, \
  and\ \bibinfo {author} {\bibfnamefont {J.}~\bibnamefont {Meng}},\ }\href
  {\doibase 10.1016/j.physletb.2006.07.064} {\bibfield  {journal} {\bibinfo
  {journal} {Phys. Lett. B}\ }\textbf {\bibinfo {volume} {640}},\ \bibinfo
  {pages} {150} (\bibinfo {year} {2006})}\BibitemShut {NoStop}%
\bibitem [{\citenamefont {Liang}\ \emph {et~al.}(2008)\citenamefont {Liang},
  \citenamefont {Van~Giai},\ and\ \citenamefont
  {Meng}}]{Liang2008Phys.Rev.Lett.101_122502}%
  \BibitemOpen
  \bibfield  {author} {\bibinfo {author} {\bibfnamefont {H.}~\bibnamefont
  {Liang}}, \bibinfo {author} {\bibfnamefont {N.}~\bibnamefont {Van~Giai}}, \
  and\ \bibinfo {author} {\bibfnamefont {J.}~\bibnamefont {Meng}},\ }\href
  {\doibase 10.1103/PhysRevLett.101.122502} {\bibfield  {journal} {\bibinfo
  {journal} {Phys. Rev. Lett.}\ }\textbf {\bibinfo {volume} {101}},\ \bibinfo
  {pages} {122502} (\bibinfo {year} {2008})}\BibitemShut {NoStop}%
\bibitem [{\citenamefont {Dirac}(1930)}]{Dirac1930Proc.Camb.Phil.Soc.26_376}%
  \BibitemOpen
  \bibfield  {author} {\bibinfo {author} {\bibfnamefont {P.~A.~M.}\
  \bibnamefont {Dirac}},\ }\href {\doibase 10.1017/S0305004100016108}
  {\bibfield  {journal} {\bibinfo  {journal} {Proc. Camb. Phil. Soc.}\ }\textbf
  {\bibinfo {volume} {26}},\ \bibinfo {pages} {376} (\bibinfo {year}
  {1930})}\BibitemShut {NoStop}%
\bibitem [{\citenamefont {Slater}(1951)}]{Slater1951Phys.Rev.81_385}%
  \BibitemOpen
  \bibfield  {author} {\bibinfo {author} {\bibfnamefont {J.~C.}\ \bibnamefont
  {Slater}},\ }\href {\doibase 10.1103/PhysRev.81.385} {\bibfield  {journal}
  {\bibinfo  {journal} {Phys. Rev.}\ }\textbf {\bibinfo {volume} {81}},\
  \bibinfo {pages} {385} (\bibinfo {year} {1951})}\BibitemShut {NoStop}%
\bibitem [{\citenamefont {Gu}\ \emph {et~al.}(2013)\citenamefont {Gu},
  \citenamefont {Liang}, \citenamefont {Long}, \citenamefont {Van~Giai},\ and\
  \citenamefont {Meng}}]{Gu2013Phys.Rev.C87_041301}%
  \BibitemOpen
  \bibfield  {author} {\bibinfo {author} {\bibfnamefont {H.-Q.}\ \bibnamefont
  {Gu}}, \bibinfo {author} {\bibfnamefont {H.}~\bibnamefont {Liang}}, \bibinfo
  {author} {\bibfnamefont {W.~H.}\ \bibnamefont {Long}}, \bibinfo {author}
  {\bibfnamefont {N.}~\bibnamefont {Van~Giai}}, \ and\ \bibinfo {author}
  {\bibfnamefont {J.}~\bibnamefont {Meng}},\ }\href {\doibase
  10.1103/PhysRevC.87.041301} {\bibfield  {journal} {\bibinfo  {journal} {Phys.
  Rev. C}\ }\textbf {\bibinfo {volume} {87}},\ \bibinfo {pages} {041301}
  (\bibinfo {year} {2013})}\BibitemShut {NoStop}%
\bibitem [{\citenamefont {Shen}\ \emph {et~al.}(2016)\citenamefont {Shen},
  \citenamefont {Hu}, \citenamefont {Liang}, \citenamefont {Meng},
  \citenamefont {Ring},\ and\ \citenamefont
  {Zhang}}]{Shen2016Chin.Phys.Lett.33_102103}%
  \BibitemOpen
  \bibfield  {author} {\bibinfo {author} {\bibfnamefont {S.-H.}\ \bibnamefont
  {Shen}}, \bibinfo {author} {\bibfnamefont {J.-N.}\ \bibnamefont {Hu}},
  \bibinfo {author} {\bibfnamefont {H.-Z.}\ \bibnamefont {Liang}}, \bibinfo
  {author} {\bibfnamefont {J.}~\bibnamefont {Meng}}, \bibinfo {author}
  {\bibfnamefont {P.}~\bibnamefont {Ring}}, \ and\ \bibinfo {author}
  {\bibfnamefont {S.-Q.}\ \bibnamefont {Zhang}},\ }\href {\doibase
  10.1088/0256-307X/33/10/102103} {\bibfield  {journal} {\bibinfo  {journal}
  {Chin. Phys. Lett.}\ }\textbf {\bibinfo {volume} {33}},\ \bibinfo {pages}
  {102103} (\bibinfo {year} {2016})}\BibitemShut {NoStop}%
\bibitem [{\citenamefont {Shen}\ \emph {et~al.}(2017)\citenamefont {Shen},
  \citenamefont {Liang}, \citenamefont {Meng}, \citenamefont {Ring},\ and\
  \citenamefont {Zhang}}]{Shen2017Phys.Rev.C96_014316}%
  \BibitemOpen
  \bibfield  {author} {\bibinfo {author} {\bibfnamefont {S.}~\bibnamefont
  {Shen}}, \bibinfo {author} {\bibfnamefont {H.}~\bibnamefont {Liang}},
  \bibinfo {author} {\bibfnamefont {J.}~\bibnamefont {Meng}}, \bibinfo {author}
  {\bibfnamefont {P.}~\bibnamefont {Ring}}, \ and\ \bibinfo {author}
  {\bibfnamefont {S.}~\bibnamefont {Zhang}},\ }\href {\doibase
  10.1103/PhysRevC.96.014316} {\bibfield  {journal} {\bibinfo  {journal} {Phys.
  Rev. C}\ }\textbf {\bibinfo {volume} {96}},\ \bibinfo {pages} {014316}
  (\bibinfo {year} {2017})}\BibitemShut {NoStop}%
\bibitem [{\citenamefont {Bender}\ \emph {et~al.}(2003)\citenamefont {Bender},
  \citenamefont {Heenen},\ and\ \citenamefont
  {Reinhard}}]{Bender2003Rev.Mod.Phys.75_121}%
  \BibitemOpen
  \bibfield  {author} {\bibinfo {author} {\bibfnamefont {M.}~\bibnamefont
  {Bender}}, \bibinfo {author} {\bibfnamefont {P.-H.}\ \bibnamefont {Heenen}},
  \ and\ \bibinfo {author} {\bibfnamefont {P.-G.}\ \bibnamefont {Reinhard}},\
  }\href {\doibase 10.1103/RevModPhys.75.121} {\bibfield  {journal} {\bibinfo
  {journal} {Rev. Mod. Phys.}\ }\textbf {\bibinfo {volume} {75}},\ \bibinfo
  {pages} {121} (\bibinfo {year} {2003})}\BibitemShut {NoStop}%
\bibitem [{\citenamefont {Liang}\ \emph {et~al.}(2015)\citenamefont {Liang},
  \citenamefont {Meng},\ and\ \citenamefont {Zhou}}]{Liang2015Phys.Rep.570_1}%
  \BibitemOpen
  \bibfield  {author} {\bibinfo {author} {\bibfnamefont {H.}~\bibnamefont
  {Liang}}, \bibinfo {author} {\bibfnamefont {J.}~\bibnamefont {Meng}}, \ and\
  \bibinfo {author} {\bibfnamefont {S.-G.}\ \bibnamefont {Zhou}},\ }\href
  {\doibase 10.1016/j.physrep.2014.12.005} {\bibfield  {journal} {\bibinfo
  {journal} {Phys. Rep.}\ }\textbf {\bibinfo {volume} {570}},\ \bibinfo {pages}
  {1} (\bibinfo {year} {2015})}\BibitemShut {NoStop}%
\bibitem [{\citenamefont {Naito}\ \emph {et~al.}(2018)\citenamefont {Naito},
  \citenamefont {Akashi},\ and\ \citenamefont
  {Liang}}]{Naito2018Phys.Rev.C97_044319}%
  \BibitemOpen
  \bibfield  {author} {\bibinfo {author} {\bibfnamefont {T.}~\bibnamefont
  {Naito}}, \bibinfo {author} {\bibfnamefont {R.}~\bibnamefont {Akashi}}, \
  and\ \bibinfo {author} {\bibfnamefont {H.}~\bibnamefont {Liang}},\ }\href
  {\doibase 10.1103/PhysRevC.97.044319} {\bibfield  {journal} {\bibinfo
  {journal} {Phys. Rev. C}\ }\textbf {\bibinfo {volume} {97}},\ \bibinfo
  {pages} {044319} (\bibinfo {year} {2018})}\BibitemShut {NoStop}%
\bibitem [{\citenamefont {Engel}\ and\ \citenamefont
  {Dreizler}(2011)}]{Engel2011_Springer-Verlag}%
  \BibitemOpen
  \bibfield  {author} {\bibinfo {author} {\bibfnamefont {E.}~\bibnamefont
  {Engel}}\ and\ \bibinfo {author} {\bibfnamefont {R.~M.}\ \bibnamefont
  {Dreizler}},\ }\href@noop {} {\emph {\bibinfo {title} {Density Functional
  Theory---An Advanced Course}}},\ Theoretical and Mathematical Physics\
  (\bibinfo  {publisher} {Springer-Verlag},\ \bibinfo {address} {Berlin,
  Heidelberg},\ \bibinfo {year} {2011})\BibitemShut {NoStop}%
\bibitem [{\citenamefont {Perdew}\ \emph
  {et~al.}(1996{\natexlab{a}})\citenamefont {Perdew}, \citenamefont {Burke},\
  and\ \citenamefont {Wang}}]{Perdew1996Phys.Rev.B54_16533}%
  \BibitemOpen
  \bibfield  {author} {\bibinfo {author} {\bibfnamefont {J.~P.}\ \bibnamefont
  {Perdew}}, \bibinfo {author} {\bibfnamefont {K.}~\bibnamefont {Burke}}, \
  and\ \bibinfo {author} {\bibfnamefont {Y.}~\bibnamefont {Wang}},\ }\href
  {\doibase 10.1103/PhysRevB.54.16533} {\bibfield  {journal} {\bibinfo
  {journal} {Phys. Rev. B}\ }\textbf {\bibinfo {volume} {54}},\ \bibinfo
  {pages} {16533} (\bibinfo {year} {1996}{\natexlab{a}})}\BibitemShut {NoStop}%
\bibitem [{\citenamefont {Perdew}\ \emph
  {et~al.}(1996{\natexlab{b}})\citenamefont {Perdew}, \citenamefont {Burke},\
  and\ \citenamefont {Ernzerhof}}]{Perdew1996Phys.Rev.Lett.77_3865}%
  \BibitemOpen
  \bibfield  {author} {\bibinfo {author} {\bibfnamefont {J.~P.}\ \bibnamefont
  {Perdew}}, \bibinfo {author} {\bibfnamefont {K.}~\bibnamefont {Burke}}, \
  and\ \bibinfo {author} {\bibfnamefont {M.}~\bibnamefont {Ernzerhof}},\ }\href
  {\doibase 10.1103/PhysRevLett.77.3865} {\bibfield  {journal} {\bibinfo
  {journal} {Phys. Rev. Lett.}\ }\textbf {\bibinfo {volume} {77}},\ \bibinfo
  {pages} {3865} (\bibinfo {year} {1996}{\natexlab{b}})}\BibitemShut {NoStop}%
\bibitem [{\citenamefont {Perdew}\ \emph {et~al.}(2008)\citenamefont {Perdew},
  \citenamefont {Ruzsinszky}, \citenamefont {Csonka}, \citenamefont {Vydrov},
  \citenamefont {Scuseria}, \citenamefont {Constantin}, \citenamefont {Zhou},\
  and\ \citenamefont {Burke}}]{Perdew2008Phys.Rev.Lett.100_136406}%
  \BibitemOpen
  \bibfield  {author} {\bibinfo {author} {\bibfnamefont {J.~P.}\ \bibnamefont
  {Perdew}}, \bibinfo {author} {\bibfnamefont {A.}~\bibnamefont {Ruzsinszky}},
  \bibinfo {author} {\bibfnamefont {G.~I.}\ \bibnamefont {Csonka}}, \bibinfo
  {author} {\bibfnamefont {O.~A.}\ \bibnamefont {Vydrov}}, \bibinfo {author}
  {\bibfnamefont {G.~E.}\ \bibnamefont {Scuseria}}, \bibinfo {author}
  {\bibfnamefont {L.~A.}\ \bibnamefont {Constantin}}, \bibinfo {author}
  {\bibfnamefont {X.}~\bibnamefont {Zhou}}, \ and\ \bibinfo {author}
  {\bibfnamefont {K.}~\bibnamefont {Burke}},\ }\href {\doibase
  10.1103/PhysRevLett.100.136406} {\bibfield  {journal} {\bibinfo  {journal}
  {Phys. Rev. Lett.}\ }\textbf {\bibinfo {volume} {100}},\ \bibinfo {pages}
  {136406} (\bibinfo {year} {2008})}\BibitemShut {NoStop}%
\bibitem [{\citenamefont {Perdew}\ \emph {et~al.}(1992)\citenamefont {Perdew},
  \citenamefont {Chevary}, \citenamefont {Vosko}, \citenamefont {Jackson},
  \citenamefont {Pederson}, \citenamefont {Singh},\ and\ \citenamefont
  {Fiolhais}}]{Perdew1992Phys.Rev.B46_6671}%
  \BibitemOpen
  \bibfield  {author} {\bibinfo {author} {\bibfnamefont {J.~P.}\ \bibnamefont
  {Perdew}}, \bibinfo {author} {\bibfnamefont {J.~A.}\ \bibnamefont {Chevary}},
  \bibinfo {author} {\bibfnamefont {S.~H.}\ \bibnamefont {Vosko}}, \bibinfo
  {author} {\bibfnamefont {K.~A.}\ \bibnamefont {Jackson}}, \bibinfo {author}
  {\bibfnamefont {M.~R.}\ \bibnamefont {Pederson}}, \bibinfo {author}
  {\bibfnamefont {D.~J.}\ \bibnamefont {Singh}}, \ and\ \bibinfo {author}
  {\bibfnamefont {C.}~\bibnamefont {Fiolhais}},\ }\href {\doibase
  10.1103/PhysRevB.46.6671} {\bibfield  {journal} {\bibinfo  {journal} {Phys.
  Rev. B}\ }\textbf {\bibinfo {volume} {46}},\ \bibinfo {pages} {6671}
  (\bibinfo {year} {1992})}\BibitemShut {NoStop}%
\bibitem [{\citenamefont {Lieb}\ and\ \citenamefont
  {Oxford}(1981)}]{Lieb1981Int.J.QuantumChem.19_427}%
  \BibitemOpen
  \bibfield  {author} {\bibinfo {author} {\bibfnamefont {E.~H.}\ \bibnamefont
  {Lieb}}\ and\ \bibinfo {author} {\bibfnamefont {S.}~\bibnamefont {Oxford}},\
  }\href {\doibase 10.1002/qua.560190306} {\bibfield  {journal} {\bibinfo
  {journal} {Int. J. Quantum Chem.}\ }\textbf {\bibinfo {volume} {19}},\
  \bibinfo {pages} {427} (\bibinfo {year} {1981})}\BibitemShut {NoStop}%
\bibitem [{\citenamefont {Lieb}\ and\ \citenamefont
  {Loss}(2001)}]{Lieb2001_AmericanMathematicalSociety}%
  \BibitemOpen
  \bibfield  {author} {\bibinfo {author} {\bibfnamefont {E.~H.}\ \bibnamefont
  {Lieb}}\ and\ \bibinfo {author} {\bibfnamefont {M.}~\bibnamefont {Loss}},\
  }\href@noop {} {\emph {\bibinfo {title} {{Analysis}}}},\ \bibinfo {edition}
  {2nd}\ ed.,\ Graduate Studies in Mathematics\ (\bibinfo  {publisher}
  {American Mathematical Society},\ \bibinfo {address} {Providence, Rhode
  Island},\ \bibinfo {year} {2001})\BibitemShut {NoStop}%
\bibitem [{\citenamefont {Levy}\ and\ \citenamefont
  {Perdew}(1985)}]{Levy1985Phys.Rev.A32_2010}%
  \BibitemOpen
  \bibfield  {author} {\bibinfo {author} {\bibfnamefont {M.}~\bibnamefont
  {Levy}}\ and\ \bibinfo {author} {\bibfnamefont {J.~P.}\ \bibnamefont
  {Perdew}},\ }\href {\doibase 10.1103/PhysRevA.32.2010} {\bibfield  {journal}
  {\bibinfo  {journal} {Phys. Rev. A}\ }\textbf {\bibinfo {volume} {32}},\
  \bibinfo {pages} {2010} (\bibinfo {year} {1985})}\BibitemShut {NoStop}%
\bibitem [{\citenamefont {Martin}(2004)}]{Martin2004_CambridgeUniversityPress}%
  \BibitemOpen
  \bibfield  {author} {\bibinfo {author} {\bibfnamefont {R.~M.}\ \bibnamefont
  {Martin}},\ }\href@noop {} {\emph {\bibinfo {title} {{Electronic Structure:
  Basic Theory and Practical Methods}}}}\ (\bibinfo  {publisher} {Cambridge
  University Press},\ \bibinfo {address} {Cambridge},\ \bibinfo {year}
  {2004})\BibitemShut {NoStop}%
\bibitem [{\citenamefont {Antoniewicz}\ and\ \citenamefont
  {Kleinman}(1985)}]{Antoniewicz1985Phys.Rev.B31_6779}%
  \BibitemOpen
  \bibfield  {author} {\bibinfo {author} {\bibfnamefont {P.~R.}\ \bibnamefont
  {Antoniewicz}}\ and\ \bibinfo {author} {\bibfnamefont {L.}~\bibnamefont
  {Kleinman}},\ }\href {\doibase 10.1103/PhysRevB.31.6779} {\bibfield
  {journal} {\bibinfo  {journal} {Phys. Rev. B}\ }\textbf {\bibinfo {volume}
  {31}},\ \bibinfo {pages} {6779} (\bibinfo {year} {1985})}\BibitemShut
  {NoStop}%
\bibitem [{\citenamefont {Vautherin}\ and\ \citenamefont
  {Brink}(1972)}]{Vautherin1972Phys.Rev.C5_626}%
  \BibitemOpen
  \bibfield  {author} {\bibinfo {author} {\bibfnamefont {D.}~\bibnamefont
  {Vautherin}}\ and\ \bibinfo {author} {\bibfnamefont {D.~M.}\ \bibnamefont
  {Brink}},\ }\href {\doibase 10.1103/PhysRevC.5.626} {\bibfield  {journal}
  {\bibinfo  {journal} {Phys. Rev. C}\ }\textbf {\bibinfo {volume} {5}},\
  \bibinfo {pages} {626} (\bibinfo {year} {1972})}\BibitemShut {NoStop}%
\bibitem [{\citenamefont {Roca-Maza}\ \emph {et~al.}(2012)\citenamefont
  {Roca-Maza}, \citenamefont {Col\`o},\ and\ \citenamefont
  {Sagawa}}]{Roca-Maza2012Phys.Rev.C86_031306}%
  \BibitemOpen
  \bibfield  {author} {\bibinfo {author} {\bibfnamefont {X.}~\bibnamefont
  {Roca-Maza}}, \bibinfo {author} {\bibfnamefont {G.}~\bibnamefont {Col\`o}}, \
  and\ \bibinfo {author} {\bibfnamefont {H.}~\bibnamefont {Sagawa}},\ }\href
  {\doibase 10.1103/PhysRevC.86.031306} {\bibfield  {journal} {\bibinfo
  {journal} {Phys. Rev. C}\ }\textbf {\bibinfo {volume} {86}},\ \bibinfo
  {pages} {031306} (\bibinfo {year} {2012})}\BibitemShut {NoStop}%
\bibitem [{\citenamefont {Col\`{o}}\ \emph {et~al.}(2013)\citenamefont
  {Col\`{o}}, \citenamefont {Cao}, \citenamefont {Van~Giai},\ and\
  \citenamefont {Capelli}}]{Colo2013Comput.Phys.Commun.184_142}%
  \BibitemOpen
  \bibfield  {author} {\bibinfo {author} {\bibfnamefont {G.}~\bibnamefont
  {Col\`{o}}}, \bibinfo {author} {\bibfnamefont {L.}~\bibnamefont {Cao}},
  \bibinfo {author} {\bibfnamefont {N.}~\bibnamefont {Van~Giai}}, \ and\
  \bibinfo {author} {\bibfnamefont {L.}~\bibnamefont {Capelli}},\ }\href
  {\doibase 10.1016/j.cpc.2012.07.016} {\bibfield  {journal} {\bibinfo
  {journal} {Comput. Phys. Commun.}\ }\textbf {\bibinfo {volume} {184}},\
  \bibinfo {pages} {142} (\bibinfo {year} {2013})}\BibitemShut {NoStop}%
\bibitem [{\citenamefont {Roca-Maza}\ and\ \citenamefont
  {Paar}(2018)}]{Roca-Maza2018Prog.Part.Nucl.Phys.101_96}%
  \BibitemOpen
  \bibfield  {author} {\bibinfo {author} {\bibfnamefont {X.}~\bibnamefont
  {Roca-Maza}}\ and\ \bibinfo {author} {\bibfnamefont {N.}~\bibnamefont
  {Paar}},\ }\href {\doibase 10.1016/j.ppnp.2018.04.001} {\bibfield  {journal}
  {\bibinfo  {journal} {Prog. Part. Nucl. Phys.}\ }\textbf {\bibinfo {volume}
  {101}},\ \bibinfo {pages} {96} (\bibinfo {year} {2018})}\BibitemShut
  {NoStop}%
\bibitem [{\citenamefont {Angeli}\ and\ \citenamefont
  {Marinova}(2013)}]{Angeli2013At.DataNucl.DataTables99_69}%
  \BibitemOpen
  \bibfield  {author} {\bibinfo {author} {\bibfnamefont {I.}~\bibnamefont
  {Angeli}}\ and\ \bibinfo {author} {\bibfnamefont {K.}~\bibnamefont
  {Marinova}},\ }\href {\doibase 10.1016/j.adt.2011.12.006} {\bibfield
  {journal} {\bibinfo  {journal} {At. Data Nucl. Data Tables}\ }\textbf
  {\bibinfo {volume} {99}},\ \bibinfo {pages} {69} (\bibinfo {year}
  {2013})}\BibitemShut {NoStop}%
\bibitem [{\citenamefont {Kim}\ \emph {et~al.}(2013)\citenamefont {Kim},
  \citenamefont {Sim},\ and\ \citenamefont
  {Burke}}]{Kim2013Phys.Rev.Lett.111_073003}%
  \BibitemOpen
  \bibfield  {author} {\bibinfo {author} {\bibfnamefont {M.-C.}\ \bibnamefont
  {Kim}}, \bibinfo {author} {\bibfnamefont {E.}~\bibnamefont {Sim}}, \ and\
  \bibinfo {author} {\bibfnamefont {K.}~\bibnamefont {Burke}},\ }\href
  {\doibase 10.1103/PhysRevLett.111.073003} {\bibfield  {journal} {\bibinfo
  {journal} {Phys. Rev. Lett.}\ }\textbf {\bibinfo {volume} {111}},\ \bibinfo
  {pages} {073003} (\bibinfo {year} {2013})}\BibitemShut {NoStop}%
\bibitem [{\citenamefont {Mohr}\ \emph {et~al.}(2016)\citenamefont {Mohr},
  \citenamefont {Newell},\ and\ \citenamefont
  {Taylor}}]{Mohr2016Rev.Mod.Phys.88_035009}%
  \BibitemOpen
  \bibfield  {author} {\bibinfo {author} {\bibfnamefont {P.~J.}\ \bibnamefont
  {Mohr}}, \bibinfo {author} {\bibfnamefont {D.~B.}\ \bibnamefont {Newell}}, \
  and\ \bibinfo {author} {\bibfnamefont {B.~N.}\ \bibnamefont {Taylor}},\
  }\href {\doibase 10.1103/RevModPhys.88.035009} {\bibfield  {journal}
  {\bibinfo  {journal} {Rev. Mod. Phys.}\ }\textbf {\bibinfo {volume} {88}},\
  \bibinfo {pages} {035009} (\bibinfo {year} {2016})}\BibitemShut {NoStop}%
\bibitem [{\citenamefont {Bulgac}\ and\ \citenamefont
  {Shaginyan}(1996)}]{Bulgac1996Nucl.Phys.A601_103}%
  \BibitemOpen
  \bibfield  {author} {\bibinfo {author} {\bibfnamefont {A.}~\bibnamefont
  {Bulgac}}\ and\ \bibinfo {author} {\bibfnamefont {V.~R.}\ \bibnamefont
  {Shaginyan}},\ }\href {\doibase 10.1016/0375-9474(96)00094-2} {\bibfield
  {journal} {\bibinfo  {journal} {Nucl. Phys. A}\ }\textbf {\bibinfo {volume}
  {601}},\ \bibinfo {pages} {103} (\bibinfo {year} {1996})}\BibitemShut
  {NoStop}%
\bibitem [{\citenamefont {Bulgac}\ and\ \citenamefont
  {Shaginyan}(1999)}]{Bulgac1999Phys.Lett.B469_1}%
  \BibitemOpen
  \bibfield  {author} {\bibinfo {author} {\bibfnamefont {A.}~\bibnamefont
  {Bulgac}}\ and\ \bibinfo {author} {\bibfnamefont {V.~R.}\ \bibnamefont
  {Shaginyan}},\ }\href {\doibase 10.1016/S0370-2693(99)01262-9} {\bibfield
  {journal} {\bibinfo  {journal} {Phys. Lett. B}\ }\textbf {\bibinfo {volume}
  {469}},\ \bibinfo {pages} {1} (\bibinfo {year} {1999})}\BibitemShut {NoStop}%
\bibitem [{\citenamefont {Carmona-Esp\'{\i}ndola}\ \emph
  {et~al.}(2015)\citenamefont {Carmona-Esp\'{\i}ndola}, \citenamefont
  {G\'{a}zquez}, \citenamefont {Vela},\ and\ \citenamefont
  {Trickey}}]{Carmona-Espindola2015J.Chem.Phys.142_054105}%
  \BibitemOpen
  \bibfield  {author} {\bibinfo {author} {\bibfnamefont {J.}~\bibnamefont
  {Carmona-Esp\'{\i}ndola}}, \bibinfo {author} {\bibfnamefont {J.~L.}\
  \bibnamefont {G\'{a}zquez}}, \bibinfo {author} {\bibfnamefont
  {A.}~\bibnamefont {Vela}}, \ and\ \bibinfo {author} {\bibfnamefont {S.~B.}\
  \bibnamefont {Trickey}},\ }\href {\doibase 10.1063/1.4906606} {\bibfield
  {journal} {\bibinfo  {journal} {J. Chem. Phys.}\ }\textbf {\bibinfo {volume}
  {142}},\ \bibinfo {pages} {054105} (\bibinfo {year} {2015})}\BibitemShut
  {NoStop}%
\end{thebibliography}%
\end{document}